%% file: ms.tex
\documentclass[runningheads]{llncs}

\usepackage[utf8x]{inputenc}
\usepackage[T1]{fontenc}
\usepackage{textcomp}
\usepackage{color}
\usepackage[lighttt]{lmodern}
\usepackage{listings}
\usepackage{graphicx}
\usepackage{hyperref}
\hypersetup{
  colorlinks = true,
  urlcolor = blue,
  linkcolor = blue,
  citecolor = blue
}

\usepackage{isabelle,isabellesym}

\isabellestyle{it}

\usepackage{ifthen}

\newcommand{\DefineSnippet}[2]{%
  \expandafter\newcommand\csname snippet--#1\endcsname{%
    \begin{quote}
    \begin{isabelle}
    \scriptsize
    #2
    \end{isabelle}
    \end{quote}}}
\newcommand{\Snippet}[1]{%
  \ifcsname snippet--#1\endcsname{\csname snippet--#1\endcsname}%
  \else+++++++ERROR: Snippet ``#1 not defined+++++++ \fi}

\input{snippets}

\newcommand{\rflx}{RecordFlux}
\newcommand{\parsers}{verifiable parsers}
\newcommand{\field}[1]{\textsf{#1}}
\newcommand{\fcode}[1]{\textit{#1}}
\newcommand{\scode}[1]{\texttt{#1}}

\begin{document}

\title{\rflx{}: Formal Message Specification and Generation of Verifiable Binary Parsers}

\titlerunning{\rflx{}}

\author{Tobias Reiher\inst{1} \and Alexander Senier\inst{1} \and Jeronimo Castrillon\inst{2} \and Thorsten Strufe\inst{2}}

\authorrunning{T. Reiher et al.}

\institute{Componolit, Dresden, Germany\\
\email{\{reiher,senier\}@componolit.com} \and
TU Dresden, Dresden, Germany\\
\email{\{jeronimo.castrillon,thorsten.strufe\}@tu-dresden.de}}

\maketitle

\begin{abstract}
Various vulnerabilities have been found in message parsers of protocol implementations in the past.
Even highly sensitive software components like TLS libraries are affected regularly.
Resulting issues range from denial-of-service attacks to the extraction of sensitive information.
The complexity of protocols and imprecise specifications in natural language are the core reasons for subtle bugs in implementations, which are hard to find.
The lack of precise specifications impedes formal verification.

In this paper, we propose a model and a corresponding domain-specific language to formally specify message formats of existing real-world binary protocols.
A unique feature of the model is the capability to define invariants, which specify relations and dependencies between message fields.
Furthermore, the model allows defining the relation of messages between different protocol layers and thus ensures correct interpretation of payload data.
We present a technique to derive verifiable parsers based on the model, generate efficient code for their implementation, and automatically prove the absence of runtime errors.
Examples of parser specifications for Ethernet and TLS demonstrate the applicability of our approach.
\end{abstract}

\section{Introduction}\label{sec:introduction}

Security issues are common in parsers of communication protocol implementations, and new vulnerabilities are found every day.
Vulnerabilities caused by incorrect parsing exist on all protocol layers: from physical and network layer protocols like Bluetooth (BlueBorne~\cite{cve:blueborne}) over session-layer protocols like TLS (Heartbleed~\cite{cve:heartbleed}) to application-layer protocols like SMB (EternalBlue~\cite{cve:eternalblue}).
Communication protocols are an increasingly worthwhile attack target, as more and more devices of our everyday life are connected to the Internet.
Their reliability is especially important in business-critical, mission-critical and safety-critical software.
Software that suddenly stops working is a potential threat to human life, be it in case of patients with an artificial heart or drivers steering their vehicles and braking using x-by-wire.
While the problem is quite obvious for highly interconnected cars, even medical devices have at least an interface for software updates, which represent an attack surface for potential compromise by targeted attacks.
Therefore, appropriate methods are needed to prevent the introduction of critical errors in protocol implementations.

Message formats of existing real-world protocols are often complex, but rarely formally specified.
The simple syntax that is commonly used only defines the basic structure of a message.
Additional properties, conditions, and relations between fields are just described in English prose.
Such descriptions are imprecise and can easily be misunderstood by developers, which leads to implementation bugs.
Lack of formal specification also prevents automatic checks and verification of the implementations.

Manual implementation has yielded 'shotgun parsers' that mix parsing and processing of messages, in the past.
The consequence have been various critical vulnerabilities~\cite{bratus2013shotgun}.
We assert that generating the parsing code from a formal grammar yields more cleanly separated implementations.

A recurrent cause of vulnerabilities is the widespread use of unsafe programming languages, like C++.
Rust and other memory-safe languages have been developed to avoid memory corruptions.
Using these languages is a clear progress towards security, but it does not prevent all errors at runtime.

Runtime errors like integer overflows or divisions by zero must still be handled explicitly.
Negligence of the matter can have devastating effects, as reported for instance in~\cite{lions1996flight}.
Formal verification is the only convincing approach towards this end.
Data and control flow analyses can prove their absence, and proving specific properties of software components is the only way to guarantee that unexpected errors do not occur at runtime.

In summary it becomes clear that a suitable process for the secure implementation of message parsers is needed.
Concluding from the observations above, we pose the following requirements.
At its heart, we need a simple, readable, and expressive domain-specific language (DSL) for a data format specification that is suitable for messages of existing real-world protocols. It shall also cover all invariants of the message parts.
It is crucial that the generated code has been verified to be free of runtime errors, to enable its use in security-critical and safety-critical applications.
To facilitate application in a wide range of areas, the generated code has to meet the performance requirements and resource limitations, even of embedded systems.

In this paper we introduce a generic approach for the specification of message formats and a methodology for creating \parsers{}.
Our main contributions are:
\begin{itemize}
    \item We propose a DSL and model for the formal specification of message formats of existing real-world binary protocols, which covers all properties and dependencies of message parts by using invariants.
    \item We introduce a methodology for the automatic generation of parsers, for which the absence of runtime errors can be shown.
    \item We show the applicability of our approach on TLS 1.3 and the TLS Heartbeat protocol.
\end{itemize}

The rest of the paper is organized as follows:
Section~\ref{sec:related_work} gives an overview of related work.
Section~\ref{sec:model} introduces the model for the specification of messages.
The design and implementation of the \rflx{} toolset is described in Section~\ref{sec:implementation}.
The applicability is shown in Section~\ref{sec:case_studies} for two case studies.
Section~\ref{sec:conclusion} gives a conclusion and an outlook for the future.

\section{Related Work}\label{sec:related_work}

In this section we describe related work for interface generators and generic parsers.

\subsubsection{Interface Generators}

Interface generators like ASN.1~\cite{asn1}, XDR~\cite{srinivasan1995xdr}, or Protocol Buffers~\cite{sw:protobuf} are used for the development of programs which communicate with each other using serialized structured data.
Although they are used to describe the message formats in various protocols or applications, they are not compatible to each other and lack the generality to specify messages of already existing protocols.
Today's commonly used communication protocols are quite complex.
Such protocols have grown historically, and therefore contain ambiguous idioms, like overlapping message fields that need to be parsed before their existence is clear.
Their representation hence is impossible with the given interface generators.

\subsubsection{Generic Parsers}

Generic parsers differ in the way how message formats are specified and which properties the generated code achieves.

One class are parser generators with a declarative description of the message structure.
PacketTypes~\cite{mccann_packet_2000} and DataScript~\cite{goos_datascript_2002} use a type-based language to describe the layout of data formats.
Binpac~\cite{pang_binpac:_2006} is a declarative language for analyzing network protocols.
GAPA~\cite{borisov_generic_2007} has a BNF-based specification language which matches the syntax commonly used in RFCs.
Kaitai Struct~\cite{sw:kaitaistruct} is YAML-based language to specify binary data formats.

Another class are parser combinators.
They combine several existing parsers that are represented by functions into a single, new parser.
Representatives are Hammer~\cite{sw:hammer}, a parsing library for binary formats written in C, attoparsec~\cite{sw:attoparsec}, a parser combinator library for Haskell, and nom~\cite{couprie2015nom}, a parser combinator written in Rust that leverages Rust's strong type system and memory safety.
Parser combinators may contain ambiguities in the grammar which are not reported at compile-time.
Consequently, parser combinators do not meet our requirements.

Parser generators in contrast can provide means to prevent ambiguities before generating code.
Many parser generators, however, use unsafe programming languages like C or C++: Binpac~\cite{pang_binpac:_2006}, PADS~\cite{daly_pads:_2006}, Nail~\cite{bangert2014nail}.
Even if the automatic nature of code generation alleviates some risk, these solutions are still prone to low-level bugs.
Some parsers like GAPA~\cite{borisov_generic_2007} especially focus on safety or use a memory-safe language.
As many errors still have to be handled correctly at runtime, these approaches are not sufficient for highly critical applications.
Lastly, parsers generated for interpreted languages~\cite{sw:construct,rodriguez2010daffodil}, which rely on a complex runtime, have limited use for resource-constraint systems.

\subsubsection{Summary}

In summary, we observe that no current solution offers expressiveness and legibility, combined with an easy venue towards formal verification and convincingly efficient generated code.
None of the analyzed approaches is expressive enough to parse binary messages of existing real-world protocols including all its properties, ensures absence of runtime errors, and is suitable for embedded systems at the same time.
Its design and implementation hence remains an open challenge.

\section{Modeling and Processing Message Formats}\label{sec:model}

In this section we introduce a methodology for the specification of message formats and subsequent generation of the corresponding \parsers{}.
Using Ethernet frames as a running example demonstrates several intricacies that require consideration.
We start with a simple variant of an Ethernet frame and refine this definition iteratively to reach a complete specification.
We set out to define the specification as a linear list of fields, like several previous approaches, but turn to a graph-based modeling later, to allow for strict specification of ambiguities in the standards.
We then describe the algorithms to generate the code of the \parsers{}.
Each parser comprises a number of functions to validate and access the content of a message.
We use Isabelle/HOL to formalize our model and describe the corresponding algorithms\footnote{Isabelle is an automated proof assistant, HOL can be used as a functional programming language that allows proving certain properties. We refer the interested reader to \cite{nipkow2002isabelle} for an introduction into the matter.}.

\subsection{Example: Ethernet Frame}

\begin{figure}[ht]
    \centering
    \includegraphics[scale=0.59]{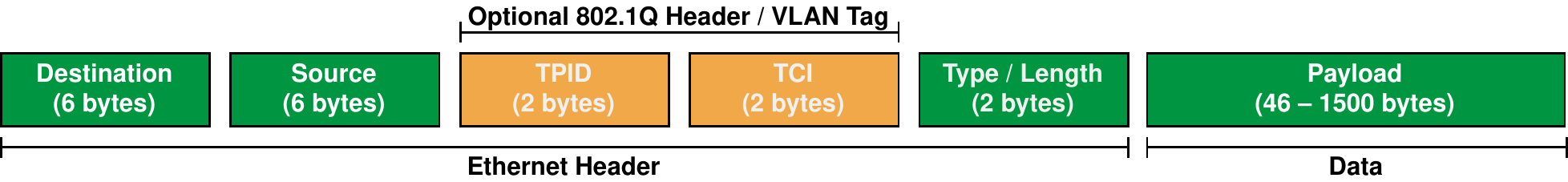}
    \caption{Ethernet frame structure}\label{fig:ethernet_frame}
\end{figure}

Figure~\ref{fig:ethernet_frame} depicts the basic structure of an Ethernet frame.
Several variants of Ethernet exist, Ethernet~II is most commonly used.
It consists of two address fields of $6$\,bytes each, a \field{Type} field of $2$\,bytes encoding which protocol is encapsulated in the payload, and a variable-length \field{Payload} field which comprises the rest of the message.
Both IEEE~802.3 and Ethernet~II frames are used in practice, so protocol instances have to distinguish them implicitly on the fly.
The payload size is limited to $1500$ bytes in Ethernet, so the field following the \field{Source} address is interpreted as IEEE~802.3 \field{Length} for values below $1500$, and Ethernet~II \field{Type} if the value is $1536$ or above.
Ethernet also defines extensions, like VLAN tagging (IEEE~802.1Q).
It inserts a VLAN tag between the \field{Source} address and the \field{Type/Length} field, which consists of two fields: the \field{TPID} field and the \field{TCI} field.
To determine if this extension is used, the instance checks the same field for the value $8100_{16}$, and interprets the bytes as the \field{TPID}, the subsequent two bytes as control information, and only the subsequent bytes as Type or Length.

\subsection{Message Representation}\label{sec:message_representation}

To generate a parser and reason about a message format like an Ethernet frame we need a formal specification that captures the message structure and all relevant constraints that must be enforced.
The simplest possible representation of a message is a list of fields.
For the Ethernet header each field can be represented by an identifier and a fixed length.
The protocol may transmit \field{Payload} of different sizes, so a variable length value is needed for this part.
To make matters worse, the payload length is defined depending on the overall length of the message in Ethernet~II, using a mathematical expression.
The underlying assumption is that there is a message buffer which comprises a number of bytes and potentially contains the message to parse.
We use a deep embedding in our model for the representation (cmp. Appendix~\ref{sec:deep_embedding}):

\Snippet{field}

Enumeration types can be used for identifiers, for instance for Ethernet~II:

\Snippet{ethernet_v2}

The message structure then is described as a list of fields.
For the \field{Destination}, \field{Source} and \field{Type} field a fixed number is sufficient to specify the field length, while for the \field{Payload} field the length of the header needs to be subtracted from the length of the message.
In our deep embedding, the variable \fcode{MessageLength} refers to the length of the message buffer.
For sake of generality we define all lengths in bits and thereby enable the definition of non-byte-granular fields.
An Ethernet~II frame is thus defined as follows:

\Snippet{ethernet_v2_frame}

To represent length fields like in the original IEEE~802.3 frame format, we need references to values of other fields in mathematical expressions.
Our deep embedding contains the constructor \fcode{FieldValue 'a} for this purpose.
Since our model works on bit granularity, but the Length field is byte-based, the value of the Length field has to be multiplied by $8$.
The specification for an IEEE~802.3 frame is as follows:

\Snippet{ethernet_frame}

To deal with the parallel use of both Ethernet frames in practice, either both formats could be treated as completely separate and handle their parallel use by other code.
We decide to combine both variations in the same model instead.
This increases model complexity slightly but prevents the need for manual handling of message variations, which could induce errors.
Combining both formats prevents representation as a linear list of fields, as the semantics of the \field{Payload Length} depends on the value of the \field{Type/Length} field.

We hence have to allow for case distinctions.
We described cases by a condition, which refers to the value of \field{Type/Length}, and a corresponding length expression for the \field{Payload} field.
In other words, there are two distinct connections between the \field{Type/Length} and the \field{Payload} field.
When modeling the dependencies between fields such connections can be interpreted as directed edges that define the order of the fields.
Such edges contain two attributes: a condition that defines when the target field follows the source field and a length expression that specifies the length of the target field.
Consequently, such edges are characterized by a source field, a target field and its attributes.
A complete message thus forms a directed-acyclic graph (DAG) where nodes represent fields.
Figure~\ref{fig:ethernet_graph} shows the graph for the full specification of Ethernet frames.

We define conditions to be handled as boolean expressions, using the same deep embedding as used for mathematical expressions.
Expressions must only contain references to field values of preceding nodes.
This restriction prevents cyclic dependencies between expressions and ensures that a sequential evaluation of the validity of a message is possible.

Allowing for VLAN tagging further complicates specification.
The existence of the 802.1Q header can only be determined by reading the two bytes following the \field{Source} address field.
To resolve the ambiguity of potential fields we add a virtual node, \fcode{Type-Length-TPID} after the \fcode{Source}.
It is solely used to differentiate the two message formats.
It is followed by both \fcode{Type-Length} as well as \fcode{TPID}.
We add a location expression at each edge that defines the position of the first bit of each respective field to be able to deal with the conditional overlay of fields in the model.
For the specification of the field location our expressions allow using \fcode{FieldFirst} and \fcode{FieldLength} to refer to the location and the length of a previous field, respectively.
We hence define edges by the following variant type, which is composed of source node, target node, condition, length expression, and location expression.

\Snippet{edge}

Finally, we want to be able to describe the restriction on the payload length, and hence the overall message length, as an invariant.
We hence introduce a final node that marks the end of the message, and as it is not followed by any other node it refers to the preceding nodes in the model, only.
We introduce an initial node that defines the beginning of a message similarly, to define the length of the first field.

At this point we are able to fully define an Ethernet frame including all its variants.
An excerpt of the full specification of Ethernet is shown below.
An unabridged version can be found in Appendix~\ref{sec:ethernet_spec}.

\Snippet{ethernet_node}

\Snippet{ethernet_graph_excerpt}

We now turn to describing the functions that are generated for the parsers corresponding to our specification.

\subsection{Derivation of Validation and Accessor Functions}

Parsers are called with a given message as input, and have to extract the content, as specified above.
They need to implement validation and accessor functions for each field, which we model as follows.
A parser~$\mathcal{P}$ consists of a list of validation and accessor functions.
The validation function allows checking if all conditions stated in the specification hold for the message field.
If this is the case, the corresponding accessor function can safely be used to retrieve the value of the field.
For each message field the code for a validation function (\fcode{FieldValidFunc}) and the respective accessor function (\fcode{FieldAccessFunc}) have to be generated.
In this manner all fields of a message can be validated and accessed consecutively.

Several conditions must hold for a bit array to be a valid message.
First of all, a message field can only be valid if its first and last bit are within the range of the message buffer.
Data out of range indicates an incomplete message.
The validity of a message field depends on the specified conditions and the validity of its predecessor.
Mapped on the model, this means that one incoming edge must have a valid condition and a valid source (except for the initial node) and, as the conditions of the outgoing edges can constrain the allowed values or the length of the field, the conditions of at least one outgoing edge must be fulfilled.
Each path from the initial node to the final node denotes a variant of a message.
A whole message is accepted if there is exactly one valid path.

Each node has to be reachable via at least one path from the initial node.
The location of a field can vary because of optional or inserted fields and thus depends on a concrete path.
Therefore the path has to be known to be able to calculate the location of a field.
As conditions can refer to other fields, the path is also needed to evaluate a condition.
For this reason, before we can validate or access a field, we have to determine all possible variants of a message, and for each variant the actual conditions and field bounds.

In the following we describe the algorithms for determining path attributes, variant functions, node paths, and field functions.
We aim for simplicity in our algorithms.
As the parser generation is only done once and the graphs which we use to represent message formats are rather small, the performance of the algorithms is not critical.

\subsubsection{Path Attributes}

As one of the first steps of the parser generation the \fcode{path-attrs} algorithm derives the attributes for all possible paths from the initial node to any node in the graph.
All references to other fields in expressions are eliminated during this process.
The starting point of the algorithm is a graph definition, where each edge can be uniquely identified by an index number, e.g., by enumerating the edges of the graph definition.

\Snippet{agraph}

The result of the algorithm is a list of tuples.
For each path, which is represented by a list of indices, an expression for the condition, the length and the location of the first bit is returned.

The algorithm iterates over all edges of the graph definition.
For each edge it determines all paths from the initial node by using the \fcode{paths} function.
Applying \fcode{concat} on the resulting list of lists gives us a list of all paths from the initial node to any other node in the graph.
Each path in this list is converted into the corresponding list of edges on the path by the \fcode{path-edges} function.
From each list of edges the last edge is taken, and for this edge the condition, length and location expression extracted.
References to other nodes in these expressions are replaced by the corresponding expression of the referenced node.
This is realized by \fcode{subs} which recursively looks up the concrete expression in the graph definition.

\Snippet{path_attrs}

\subsubsection{Variant Functions}

The parser~$\mathcal{P}$ contains a variant validation function \fcode{VariantValidFunc} and a variant accessor function \fcode{VariantAccessFunc} for each message variant to allow the validation and access of a concrete variant of a field.
These variant functions form the building blocks of the field validation functions and the field accessor functions.
For each tuple generated by \fcode{path-attrs} containing condition, length and location for a specific path, a \fcode{VariantValidFunc} and a \fcode{VariantAccessFunc} are derived.

The body of a \fcode{VariantValidFunc} is based on the condition, a check which ensures that the field is within the bounds of the input buffer, and a call to the validation function of the preceding field, if it is not the first field of the message.
Each variant function is identified by a path.
As a path is represented by a list of indices, the preceding field can be determined by removing the last element of the current path.\footnote{The \fcode{init} function returns a list without its last element.}
Calls to other variant functions are denoted by \fcode{VariantValidCall} and \fcode{VariantAccessCall}, respectively.

\Snippet{variant_valid_funcs}

A \fcode{VariantAccessFunc} is defined by the location expression and the length expression of the field.

\Snippet{variant_access_funcs}

The result of \fcode{variant-valid-funcs} and \fcode{variant-access-funcs} form the variant functions $\mathcal{V}$.

\subsubsection{Node Paths}

The \fcode{node-paths} algorithm determines which paths lead to a field, i.e., which variants of a field exist, and which conditions at outgoing edges a node has.
As described in Section~\ref{sec:message_representation} the values of a field can be further restricted by outgoing edges.
At least one condition of an outgoing edge has to be fulfilled.
Therefore, the corresponding conditions have to be determined as well.
Like before, all references to other fields need to be resolved in dependence of a variant.

\fcode{node-paths} iterates over all nodes of the graph.
The list of nodes of a graph is provided by \fcode{graph-nodes}.
For each node all incoming edges are determined by \fcode{incoming}.
Each incoming edge is used to determine all paths from the initial node by \fcode{paths}.
For each path it then creates a tuple with two elements: the path represented by a list of indices and a disjunction of all conditions at outgoing edges.
The disjunction is created by \fcode{any}, which takes a list of conditions from \fcode{path-conds}.
\fcode{path-conds} extracts the conditions of the list of outgoing edges determined by \fcode{path-edges} and \fcode{outgoing}.
Finally, the list of tuples is assigned to the corresponding node identifier.

\Snippet{node_paths}

\subsubsection{Field Functions}

A \fcode{FieldValidFunc} determines if one variant is valid.
If a valid variant exists, the \fcode{FieldAccessFunc} can be used to return the value of the field.
Field functions rely on the functionality provided by variant functions.

The resulting list of \fcode{node-paths} is used to generate validation and accessor functions for each field of the message.
Each element of this list contains all the necessary information to create a validation and accessor function for one field.
A field function is identified by a node identifier.

The algorithm \fcode{field-valid-funcs} creates a list of \fcode{FieldValidFunc}.
In order that a field is valid, a variant of the field and the conditions of one of the outgoing edges must be valid.
Hence, the body of a \fcode{FieldValidFunc} is a disjunction of calls to all variant validation functions and the corresponding expression which was determined for the conditions at outgoing edges.

\Snippet{field_valid_funcs}

The body of each function is determined by \fcode{valid-calls}.
\fcode{valid-calls} iterates over all path-condition tuples which it receives as arguments.
For each path it creates a call to a \fcode{VariantValidFunc} and combines this call with the corresponding expression derived from the outgoing edges by a conjunction, as a variant is only valid if one of the conditions at the outgoing edges is valid.
All created conjunctions are connected by a disjunction, as only one variant has to be valid.

\Snippet{valid_calls}

The list of field accessor functions is created by \fcode{field-access-funcs}.
A \fcode{FieldAccessFunc} checks subsequently which variant of a field is valid and calls the corresponding \fcode{VariantAccessFunc}.

\Snippet{field_access_funcs}

The body of such a function is created by \fcode{access-calls}.
It iterates over the list of paths and creates a nested if-expression, where the else-branch is created recursively.
Each if-expression has a call to a \fcode{VariantValidFunc} as condition and a call to the corresponding \fcode{VariantAccessFunc} as body.

\Snippet{access_calls}

The result of \fcode{field-valid-funcs} and \fcode{field-access-funcs} form the field functions~$\mathcal{F}$.
The parser~$\mathcal{P}$ comprises all variant functions~$\mathcal{V}$ and field functions~$\mathcal{F}$.

\subsection{Message Refinement}

Communication protocols are typically structured in layers.
A protocol message contains a message of a higher layer protocol as its payload.
We model this relation by message refinements.
A message refinement is a tuple consisting of an identifier of the message, the name of the payload field, an identifier for the contained message and an expression.
The expression describes under which conditions a message is contained in the payload field of another message.
In the case of Ethernet the expression could specify that an IPv4 packet is contained in the Ethernet frame's payload field, if the \field{Type/Length} field has the value $0800_{16}$.
For each message multiple message refinements can be defined.

\section{Implementation}\label{sec:implementation}

\begin{figure}
    \centering
    \includegraphics[scale=0.59]{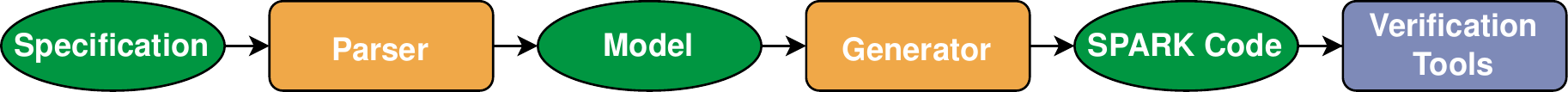}
    \caption{Architecture}\label{fig:architecture}
\end{figure}

The \rflx{} toolset\footnote{\rflx{} is available as open source~\cite{sw:recordflux}.} comprises multiple parts (Figure~\ref{fig:architecture}).
The specification language allows describing message formats and the relation of a message field to messages of higher protocol layers.
The specification parser transforms this textual description into the model introduced in Section~\ref{sec:model}, which is used by the code generator.
We chose SPARK~\cite{sw:spark} as the target language for code generation, as it already provides simple verification including all required tools.
It is supported by the standard GCC toolchain and suitable for resource constrained systems.\footnote{We refer the interested reader \cite{mccormick2015building} for an introduction into the language including all of its beneficial properties.}
We hence generate SPARK code, including all necessary function contracts.
We then use the SPARK verification toolset to ensure the absence of runtime errors and the functional correctness of the generated code.

\subsection{Specification Language}\label{sec:specification_language}

To specify message formats in a simple and readable manner, we have designed a specification language that allows expressing all properties of a message in accordance to our model.
The specification language describes messages based on types.
A type definition has the form: \scode{\textbf{type} NAME \textbf{is} DEFINITION;}

The language supports two integer types to represent numbers: modular and range integers.
A modular type represents the values from zero to one less than the modulus.
The bit size of a modular type is determined by calculating the binary logarithm of the modulus.
The destination and source address fields of Ethernet is represented by the following modular integer:

\begin{lstlisting}
   type Address is mod 2**48;
\end{lstlisting}

A range integer allows restricting the range of numbers by bounds.
The set of values of a range type consists of all numbers from the lower bound to the upper bound.
For a range type the bit size has to be specified explicitly.
A range integer can be used for the \field{Type/Length} field, and allows incorporating the minimum length restriction of the payload field into the type definition:

\begin{lstlisting}
   type Type_Length is range 46 .. 2**16 - 1 with Size => 16;
\end{lstlisting}

This defines a type with a size of $16$ bit which comprises all numbers from $46$ to $2^{16} - 1$.

A message format is specified by a message type.
A message type is a collection of components.
Each component corresponds to one field in a message and is of form: \scode{FIELD\_NAME : FIELD\_TYPE}.
A simplified specification of an Ethernet~II frame is as follows:

\begin{lstlisting}
   type Simplified_Frame is
      message
         Destination : Address;
         Source : Address;
         Type_Length : Type_Length;
         Payload : Payload;
      end message;
\end{lstlisting}

But as argued in Section~\ref{sec:model} such a simple specification is not sufficient for Ethernet in general.

\begin{figure}[ht]
    \begin{minipage}[b]{0.48\textwidth}
        \begin{lstlisting}[basicstyle=\ttfamily\tiny]
package Ethernet is

   type Address is mod 2**48;
   type Type_Length is range 46 .. 2**16 - 1
      with Size => 16;
   type TPID is range 16#8100# .. 16#8100#
      with Size => 16;
   type TCI is mod 2**16;

   type Frame is
     message
       Destination : Address;
       Source : Address;
       Type_Length_TPID : Type_Length
          then TPID
             with First => Type_Length_TPID'First
             if Type_Length_TPID = 16#8100#,
          then Type_Length
             with First => Type_Length_TPID'First
             if Type_Length_TPID /= 16#8100#;
       TPID : TPID;
       TCI : TCI;
       Type_Length : Type_Length
          then Payload
             with Length => Type_Length * 8
             if Type_Length <= 1500,
          then Payload
             with Length => Message'Last - Type_Length'Last
             if Type_Length >= 1536;
       Payload : Payload
          then null
             if Payload'Length / 8 >= 46
                 and Payload'Length / 8 <= 1500;
     end message;

end Ethernet;
        \end{lstlisting}
        \caption{Full specification of an Ethernet frame covering Ethernet~II, IEEE~802.3, and IEEE~802.1Q \hspace{.1\linewidth}\vspace{1\baselineskip}}\label{fig:ethernet_spec}
    \end{minipage}
    \hspace{.02\linewidth}
    \begin{minipage}[b]{0.48\textwidth}
        \centering
        \includegraphics[scale=0.55]{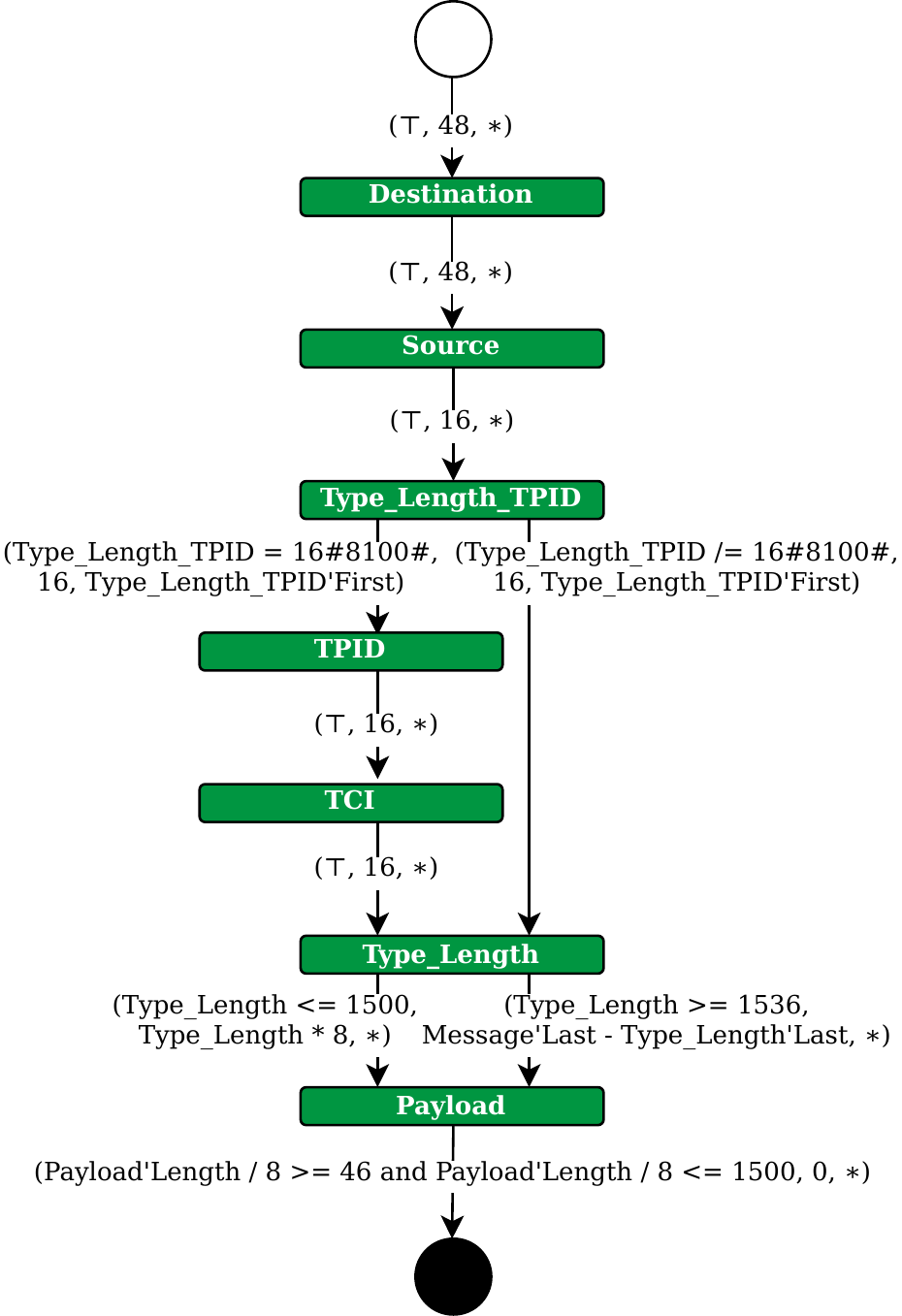}
        \vspace{\baselineskip}
        \caption{Graph representation of Ethernet frame specification (Notation: For an edge e = (s,t,c,l,f): $\ast$ denotes \scode{f = s'First + s'Length}, $\top$ denotes \scode{c = True})}\label{fig:ethernet_graph}
    \end{minipage}
\end{figure}

A then clause following a component allows defining which field follows.
If no then clause is given, it is assumed that always the next component of the message follows.
If no further component follows, it is assumed that the message ends with this field.
A then clause can contain a condition under which the corresponding field follows and aspects which allow defining the length of the next field and the location of its first bit.
The condition can refer to previous fields (including the component containing the then clause).
In case of Ethernet two then clauses can be added to the \field{Type/Length} field to differentiate the two different meanings of this field:

\begin{lstlisting}
      Type_Length : Type_Length
         then Payload
            with Length => Type_Length * 8
            if Type_Length <= 1500,
         then Payload
            with Length => Message'Last - Type_Length'Last
            if Type_Length >= 1536;
\end{lstlisting}

The full specification of an Ethernet frame including VLAN tags is shown in Figure~\ref{fig:ethernet_spec}.
Figure~\ref{fig:ethernet_graph} depicts the corresponding graph representation.
The package \scode{Ethernet} consists of multiple integer types and a message type \scode{Frame}.
Packages are used to structure a specification and thus make the specification modular.

A type refinement describes the relation of a component in a message type to another message type.
It states under which condition a specific protocol message is expected inside of a payload field.
Only components of the built-in type \scode{Payload} can be refined.
Types defined in other packages are referenced by a qualified name in the form \scode{package\_name.message\_type\_name}.
The condition can refer to components of the refined type.

\begin{lstlisting}
   type IPv4_In_Ethernet is new Ethernet.Frame (Payload => IPv4.Packet)
      if Type_Length = 16#0800#;
\end{lstlisting}

In this example the relation between an Ethernet frame and an IPv4 packet is specified.
The message type \scode{Frame} in package \scode{Ethernet} contains a \scode{Packet} defined in package \scode{IPv4} if the \scode{Type\_Length} field of the Ethernet frame equals to \scode{0x0800}.

\subsection{Code Generation}\label{sec:code_generation}

The basis for the code generation is the model described in Section~\ref{sec:model}.
The generated code takes a plain byte array as input and allows validating and accessing the message data in a structured way.
For each specified message a number of functions is generated.
The user of the generated code finds a validation function and accessor function for each field of the message.
The validity of a field must be checked before accessing its value.
This is realized by preconditions.
By this means it is ensured that the value of a field is only accessible if all previous fields and the value of the field is valid.

Applying a function to a wrong input buffer or to an incorrect part of the buffer could lead to unexpected results.
To prevent this a buffer has to be labeled correctly.
This is realized by a predicate used as precondition of all validation and accessor functions.
A label is added automatically if the relation between a payload field and a contained message is specified by a type refinement, and a contains function is used to check if the corresponding conditions are fulfilled for the input buffer in question.
A contains function is the representation of a type refinement in the generated code.
If the input data is received from an external source, the input buffer must be labeled explicitly.

The structure of the specification is reflected in the generated code.
As a result it is possible to keep the code as well as the specification modular and extendable.
For example type refinements can be defined in a separate specification.
This allows adding further higher layer protocols transmitted in an already specified protocol without changing existing code.

\subsection{Verification}\label{sec:verification}

The SPARK programming language allows the detailed specification of the behavior of software components by the use of contracts.
This specification is used by the SPARK verification tools to formally proof that the stated properties of the program hold.
The achievable assurance ranges from showing that no runtime exceptions occur to ensuring functional correctness based on a formal specification.
This is realized by analyzing the source code and generating verification conditions, which are then passed to multiple theorem provers to formally verify the correctness of the code.

The use of SPARK allows us proving the absence of runtime errors and the correct use of the generated code.
All of the generated code is valid SPARK code and will be analyzed by the verification tools.
The incorrect use of the generated code, e.g.~accessing a field value without prior verification, is prevented by adding appropriate contracts to these functions.

A key benefit of using SPARK is that the code generator need not to be trusted with regard to the absence of runtime errors in the generated code, as this property is proved by the verification tools.
Furthermore, the SPARK verification tools assist ensuring the correctness of the specified message format.
For example the tools will find potential integer overflows in expressions, which could indicate a missing restriction of the value range of a field.

\section{Case Studies}\label{sec:case_studies}

We demonstrate the applicability of \rflx{} using the example of TLS in two case studies.
In the first study we replaced the whole parsing code of an existing TLS library by an implementation specified and generated by \rflx{} and analyzed its impact.
In the second study we used \rflx{} to specify and parse messages of the TLS Heartbeat protocol, an optional extension of the TLS standard, which is not supported by the library used in the first study.

\subsection{Verified TLS Parser}

Fizz~\cite{nekritz2018fizz} is a TLS 1.3 implementation developed and used by Facebook, written in C++.
As a proof of concept we have replaced the C++ parser by verified SPARK code.
Therefore, we used our specification language (see Section~\ref{sec:specification_language}) to specify the messages of TLS 1.3, as standardized in RFC8446.
Based on this specification, \rflx{} generated SPARK code for TLS Record messages, TLS Handshake messages and TLS extensions.
We integrated the existing C++ code with the generated SPARK code manually.
The glue code mainly performs the conversion between C++-specific structures like vectors and SPARK-compatible data formats.

\subsubsection{Security}

Parsing of protocol messages is a sensitive part of a TLS implementation.
\cite{backhouse2019facebook} reports an integer overflow in Fizz.
An exploit could have left the application using Fizz in an infinite loop, just by sending a short sequence of messages with a well-chosen value in the length field of a TLS Record message.
Facebook fixed the bug by choosing a bigger integer type (\scode{size\_t} instead of \scode{uint16}).
\rflx{} checks the length field for allowed values before continuing the parsing, which we argue is a better solution to the problem.
The SPARK verification tools can then prove the absence of unexpected integer overflows.

\subsubsection{Performance}

\begin{figure}[ht]
    \begin{minipage}{0.48\textwidth}
        \centering
        \includegraphics[width=\textwidth]{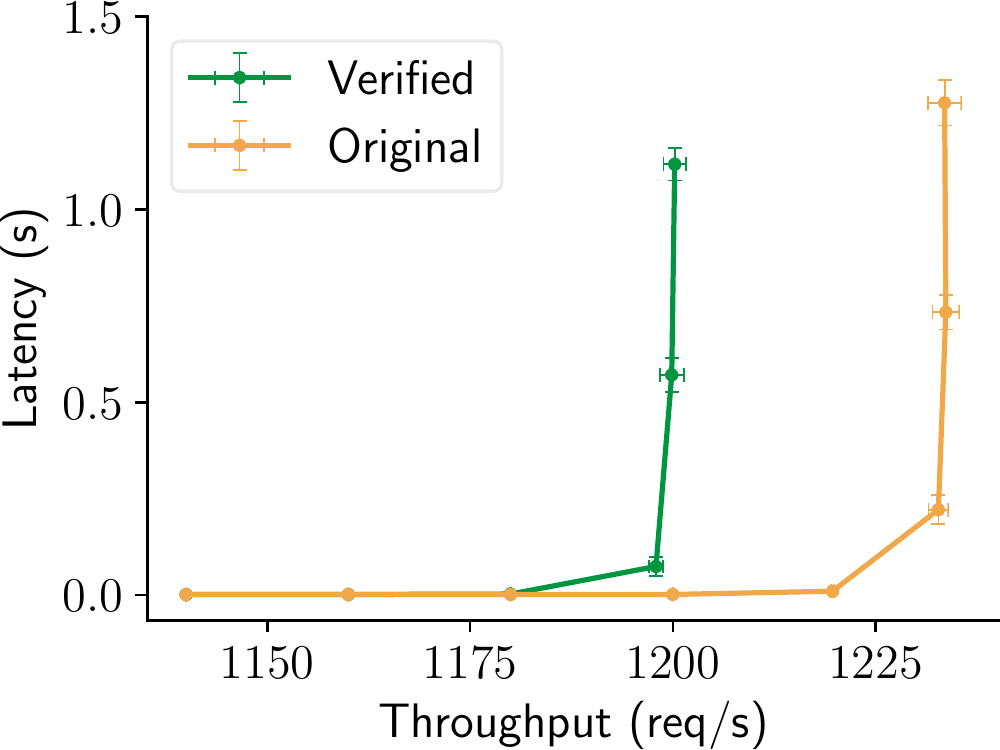}
        \caption{Performance at Handshake layer; separate TLS handshake for each request}\label{fig:fizz_perfomance_handshake}
    \end{minipage}
    \hspace{.02\linewidth}
    \begin{minipage}{0.48\textwidth}
        \centering
        \includegraphics[width=\textwidth]{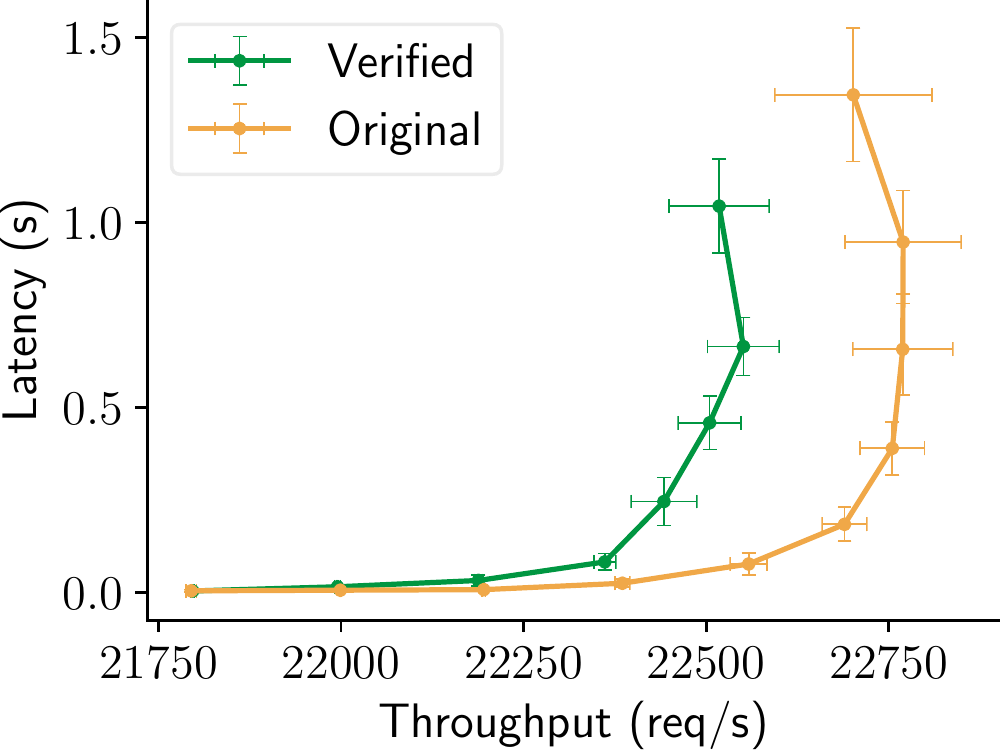}
        \caption{Performance at Record layer; one TLS handshake for multiple requests}\label{fig:fizz_perfomance_record}
    \end{minipage}
\end{figure}

Performance is considered at least of equal importance as security in practice.
We hence evaluated the performance impact of replacing the original message parser with the code generated by \rflx.
For that purpose we used a modified version of wrk2~\cite{sw:wrk2_mod}, where we added the possibility to run in two different modes.
To measure the impact of \rflx{} on the TLS handshake, the first mode of wrk2 creates a TLS connection for each HTTP request.
The impact of \rflx{} during data transmission is measured by the second mode of wrk2 that only creates one TLS connection before sending requests.
wrk2 sends requests in a constant rate and measures the resulting throughput and latency of the responses.
The sending rate of wrk2 is increased iteratively until the throughput is not improving any more.
For each sending rate the mean values of 40~measurements with a duration of 60~seconds each are calculated.
To minimize the impact of network hardware on the results we run Fizz and wrk2 on the same machine.

We expected to see some performance impact due to the additional validation checks in the generated code and the conversions between C++ and SPARK structures.
The diagrams in Figures~\ref{fig:fizz_perfomance_handshake} and~\ref{fig:fizz_perfomance_record} show the resulting mean values of throughput and latency and the corresponding 95\,\% confidence intervals.
The maximum throughput is around 2.7\,\% lower in the Handshake layer and 1.1\,\% lower in the Record layer compared to the original parser.
An analysis of the CPU cycles used by both variants with Valgrind~\cite{nethercote2007valgrind} showed that the majority of additional cycles are spent on memory allocations and processing of data conversions in the glue code.

The results show that there is no significant performance degradation.
We conclude that the approach is generally applicable, although mixing existing C++ code with SPARK code is not ideal from the point of view of performance.

\subsection{TLS Heartbeat}

\begin{figure}[ht]
    \centering
    \includegraphics[scale=0.59]{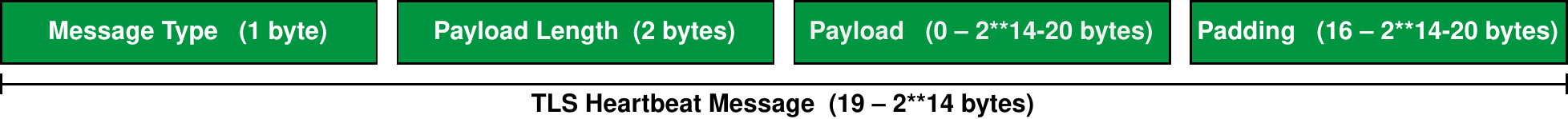}
    \caption{Message format of a TLS Heartbeat}\label{fig:heartbeat_message}
\end{figure}

The Heartbeat extension adds keep-alive functionality to TLS\@.
It gained inglorious prominence by Heartbleed~\cite{cve:heartbleed}, a security vulnerability in the OpenSSL library that affected millions of devices.
Heartbleed allowed to extract sensitive data from a TLS endpoint due to an improper input validation.

Both sides of a TLS connection can request the use of the Heartbeat protocol during the TLS handshake.
If accepted by the other side, the initiator is allowed to periodically send Heartbeat requests during the lifetime of the TLS connection.
Each request contains payload of arbitrary length and content.
The receiver of a Heartbeat request must send a response back which contains the same payload as the request.
The format of a TLS Heartbeat message is shown in Figure~\ref{fig:heartbeat_message}.
The corresponding \rflx{} specification of a TLS Heartbeat message is as follows:

\begin{lstlisting}
package TLS_Heartbeat is

   type Message_Type is (HEARTBEAT_REQUEST => 1, HEARTBEAT_RESPONSE => 2) with Size => 8;
   type Length is range 0 .. 2**14 - 20 with Size => 16;

   type Heartbeat_Message is
      message
         Message_Type : Message_Type;
         Payload_Length : Length
            then Payload with Length = Payload_Length * 8;
         Payload : Payload
            then Padding with Length = Message'Last - Payload'Last;
         Padding : Payload
            then null if Message'Length <= 2**14 * 8 and Padding'Length >= 16 * 8;
      end message;

end TLS_Heartbeat;
\end{lstlisting}

\scode{Heartbeat\_Message} represents a TLS Heartbeat Message.
Such a message consists of four fields:
The \scode{Message\_Type} field specifies the type of the message.
It is represented by a enumeration type with a size of 1~byte and comprises two valid values:
\scode{1} for a request and \scode{2} for a response.
All other values are considered invalid.
\scode{Payload\_Length} defines the length of the following \scode{Payload} field.
The \scode{Payload} field contains the content of the message, and \scode{Padding} comprises the rest of the message.

The lengths of the \scode{Payload} and the \scode{Padding} field is explicitly defined by length expressions.
The whole message is restricted to a length of $2^{14}$~bytes.
The \scode{Padding} field must be at least $16$~bytes long.

The following excerpt shows some of the generated SPARK subprogram declarations for the a \scode{Heartbeat\_Message}:

\begin{lstlisting}
function Is_Contained (Buffer : Bytes) return Boolean with Ghost, Import;

procedure Label (Buffer : Bytes) with Ghost, Post => Is_Contained (Buffer);

function Valid_Message_Type (Buffer : Bytes) return Boolean
  with Pre => Is_Contained (Buffer);

function Get_Message_Type (Buffer : Bytes) return Message_Type
  with Pre => (Is_Contained (Buffer) and then Valid_Message_Type (Buffer));

function Valid_Payload (Buffer : Bytes) return Boolean
  with Pre => Is_Contained (Buffer);

procedure Get_Payload (Buffer : Bytes; First : out Natural; Last : out Natural)
  with Pre => (Is_Contained (Buffer) and then Valid_Payload (Buffer)),
       Post => (First = Get_Payload_First (Buffer) and then
                Last = Get_Payload_Last (Buffer));

function Is_Valid (Buffer : Bytes) return Boolean
  with Pre => Is_Contained (Buffer);
\end{lstlisting}

The \scode{Is\_Contained} function which is a precondition for all validation and accessor functions ensures that always the correct message buffer is used.
The \scode{Is\_Contained} predicate is set automatically for all defined message refinements.
If a message buffer is received from an external source it can be explicitly labeled with the \scode{Label} function.
For each message field a validation function (prefixed with \scode{Valid}) and an accessor function (prefixed with \scode{Get}) is created.
Each accessor function has the corresponding validation function as a precondition.
This ensures that the validity is always checked before accessing a field value.
The \scode{Is\_Valid} checks the validity of the whole message.
It returns \scode{True} if the input buffer contains one valid message variant.

The following code example shows how the generated code is used:

\begin{lstlisting}
with IO;
with TLS.Heartbeat_Message; use TLS.Heartbeat_Message;

procedure Main is
   Buffer : Bytes := IO.Read;
   Tag    : Message_Type;
   First  : Natural;
   Last   : Natural;
begin
   Label (Buffer);
   if Is_Valid (Buffer) then
      Tag := Get_Message_Type (Buffer);
      Get_Payload (Buffer, First, Last);
      Process_Payload (Buffer (First .. Last));
   end if;
end Main;
\end{lstlisting}

The message buffer is read from an external source.
In this example the message buffer is explicitly labeled as a buffer which should contain a TLS Heartbeat message.
By also specifying the Record layer with \rflx{} and defining a message refinement between Record message and Heartbeat message, no labeling would be needed.
The validity of content of \scode{Buffer} is checked by \scode{Is\_Valid}.
Alternatively a user could also check the validity of each field on his own.
As the \scode{Padding} must be ignored and the relation between the \scode{Payload\_Length} field and the \scode{Payload} field is internally known, a user only needs to access the \scode{Message\_Type} and the \scode{Payload}.

The SPARK verification tools ensure the correct use of the generated code.
If a user would not check the validity of the input buffer, the tools will find this mistake when proving the correctness of the code.
\scode{Get\_Message\_Type} and \scode{Get\_Payload} will be flagged with \scode{precondition might fail}.

While the Heartbeat protocol appears quite simple, a flawed implementation can have serious implications.
Heartbleed allowed to send a request with a high length value while sending just a short payload and padding.
On the receiver side the length value was not checked against the actual received payload.
This led to a buffer overflow, so that not only the payload of the request was sent back, but also data following the message buffer.

\rflx{} enforces that the length of a payload field is always defined by a length expression.
The code generator adds checks to ensure that the value of a length field is in the allowed range and the message to parse is long enough, and so prevents the issue seen in Heartbleed.
Even if \rflx{} would erroneously miss adding a necessary check, the SPARK verification tools will find the potential buffer-overflow before faulty code is used involuntarily.

\section{Conclusion and Outlook}\label{sec:conclusion}

We have created a methodology for specification of message formats of communication protocols and the automatic generation of a parser.
Based on this methodology we have created a practical implementation, which comprises a DSL for describing message formats and code generator that creates SPARK code, for which the absence of runtime errors can be shown.
The generated code is applicable in real-world applications as demonstrated for TLS 1.3, which proved to suffer only from minor performance penalties despite its proven security.

So far we only handled the parsing of messages, but this is only one part of a protocol.
In the future we will also look into the protocol logic.
We aim to extend the current methodology to get a full formal specification of a protocol, to generate provable secure code.

\subsubsection{Acknowledgements}

This work is partially funded by the European Union (EU), the European Social Fund (ESF), tax money on the basis of the budget approved by the members of the parliament of Saxony, and the Cluster of Excellence EXC 2050 ``Centre for Tactile Internet with Human-in-the-Loop'' (CeTI).

\hbadness=10000
\bibliographystyle{splncs04}
\bibliography{bibliography}

\newpage
\appendix
\section{Deep Embedding}\label{sec:deep_embedding}

\Snippet{expr}

\section{Formal Specification of Ethernet Frame}\label{sec:ethernet_spec}

\Snippet{ethernet_graph}

\end{document}

%% file: snippets.tex
\DefineSnippet{init}{
\isacommand{definition}\isamarkupfalse%
\ init\ {\isacharcolon}{\isacharcolon}\ {\isachardoublequoteopen}{\isacharprime}a\ list\ {\isasymRightarrow}\ {\isacharprime}a\ list{\isachardoublequoteclose}\ \isakeyword{where}\isanewline
{\isachardoublequoteopen}init\ l\ {\isacharequal}\ {\isacharparenleft}rev\ {\isacharparenleft}drop\ {\isadigit{1}}\ {\isacharparenleft}rev\ l{\isacharparenright}{\isacharparenright}{\isacharparenright}{\isachardoublequoteclose}%
}
\DefineSnippet{enumerate}{
\isacommand{definition}\isamarkupfalse%
\ enumerate\ {\isacharcolon}{\isacharcolon}\ {\isachardoublequoteopen}{\isacharprime}a\ list\ {\isasymRightarrow}\ {\isacharparenleft}nat\ {\isasymtimes}\ {\isacharprime}a{\isacharparenright}\ list{\isachardoublequoteclose}\ \isakeyword{where}\isanewline
{\isachardoublequoteopen}enumerate\ l\ {\isacharequal}\ {\isacharbrackleft}{\isacharparenleft}i{\isacharcomma}\ l\ {\isacharbang}\ i{\isacharparenright}\ {\isachardot}\ i\ {\isasymleftarrow}\ upt\ {\isadigit{0}}\ {\isacharparenleft}length\ l{\isacharparenright}{\isacharbrackright}{\isachardoublequoteclose}%
}
\DefineSnippet{expr}{
\isacommand{datatype}\isamarkupfalse%
\ {\isacharprime}a\ expr\ {\isacharequal}\isanewline
\ \ Num\ int\ {\isacharbar}\isanewline
\ \ Add\ {\isachardoublequoteopen}{\isacharprime}a\ expr{\isachardoublequoteclose}\ {\isachardoublequoteopen}{\isacharprime}a\ expr{\isachardoublequoteclose}\ {\isacharbar}\isanewline
\ \ Mul\ {\isachardoublequoteopen}{\isacharprime}a\ expr{\isachardoublequoteclose}\ {\isachardoublequoteopen}{\isacharprime}a\ expr{\isachardoublequoteclose}\ {\isacharbar}\isanewline
\ \ Sub\ {\isachardoublequoteopen}{\isacharprime}a\ expr{\isachardoublequoteclose}\ {\isachardoublequoteopen}{\isacharprime}a\ expr{\isachardoublequoteclose}\ {\isacharbar}\isanewline
\ \ Div\ {\isachardoublequoteopen}{\isacharprime}a\ expr{\isachardoublequoteclose}\ {\isachardoublequoteopen}{\isacharprime}a\ expr{\isachardoublequoteclose}\ {\isacharbar}\isanewline
\ \ Value\ {\isachardoublequoteopen}{\isacharprime}a\ expr{\isachardoublequoteclose}\ {\isachardoublequoteopen}{\isacharprime}a\ expr{\isachardoublequoteclose}\ {\isacharbar}\isanewline
\ \ FieldValue\ {\isacharprime}a\ {\isacharbar}\isanewline
\ \ FieldLength\ {\isacharprime}a\ {\isacharbar}\isanewline
\ \ FieldFirst\ {\isacharprime}a\ {\isacharbar}\isanewline
\ \ MessageLength\ {\isacharbar}\isanewline
\ \ MessageFirst\ {\isacharbar}\isanewline
\ \ MessageLast\ {\isacharbar}\isanewline
\ \ BufferLength\ {\isacharbar}\isanewline
\ \ VariantValidCall\ {\isachardoublequoteopen}nat\ list{\isachardoublequoteclose}\ {\isacharbar}\isanewline
\ \ VariantAccessCall\ {\isachardoublequoteopen}nat\ list{\isachardoublequoteclose}\ {\isacharbar}\isanewline
\ \ Null\ {\isacharbar}\isanewline
\ \ True\ {\isacharbar}\isanewline
\ \ And\ {\isachardoublequoteopen}{\isacharprime}a\ expr{\isachardoublequoteclose}\ {\isachardoublequoteopen}{\isacharprime}a\ expr{\isachardoublequoteclose}\ {\isacharbar}\isanewline
\ \ Or\ {\isachardoublequoteopen}{\isacharprime}a\ expr{\isachardoublequoteclose}\ {\isachardoublequoteopen}{\isacharprime}a\ expr{\isachardoublequoteclose}\ {\isacharbar}\isanewline
\ \ Eq\ {\isachardoublequoteopen}{\isacharprime}a\ expr{\isachardoublequoteclose}\ {\isachardoublequoteopen}{\isacharprime}a\ expr{\isachardoublequoteclose}\ {\isacharbar}\isanewline
\ \ Ne\ {\isachardoublequoteopen}{\isacharprime}a\ expr{\isachardoublequoteclose}\ {\isachardoublequoteopen}{\isacharprime}a\ expr{\isachardoublequoteclose}\ {\isacharbar}\isanewline
\ \ Lt\ {\isachardoublequoteopen}{\isacharprime}a\ expr{\isachardoublequoteclose}\ {\isachardoublequoteopen}{\isacharprime}a\ expr{\isachardoublequoteclose}\ {\isacharbar}\isanewline
\ \ Le\ {\isachardoublequoteopen}{\isacharprime}a\ expr{\isachardoublequoteclose}\ {\isachardoublequoteopen}{\isacharprime}a\ expr{\isachardoublequoteclose}\ {\isacharbar}\isanewline
\ \ Gt\ {\isachardoublequoteopen}{\isacharprime}a\ expr{\isachardoublequoteclose}\ {\isachardoublequoteopen}{\isacharprime}a\ expr{\isachardoublequoteclose}\ {\isacharbar}\isanewline
\ \ Ge\ {\isachardoublequoteopen}{\isacharprime}a\ expr{\isachardoublequoteclose}\ {\isachardoublequoteopen}{\isacharprime}a\ expr{\isachardoublequoteclose}\ {\isacharbar}\isanewline
\ \ IfThenElse\ {\isachardoublequoteopen}{\isacharprime}a\ expr{\isachardoublequoteclose}\ {\isachardoublequoteopen}{\isacharprime}a\ expr{\isachardoublequoteclose}\ {\isachardoublequoteopen}{\isacharprime}a\ expr{\isachardoublequoteclose}%
}
\DefineSnippet{edge}{
\isacommand{datatype}\isamarkupfalse%
\ {\isacharprime}a\ edge\ {\isacharequal}\isanewline
\ \ Edge\ {\isachardoublequoteopen}{\isacharprime}a{\isachardoublequoteclose}\ {\isachardoublequoteopen}{\isacharprime}a{\isachardoublequoteclose}\ {\isachardoublequoteopen}{\isacharprime}a\ expr{\isachardoublequoteclose}\ {\isachardoublequoteopen}{\isacharprime}a\ expr{\isachardoublequoteclose}\ {\isachardoublequoteopen}{\isacharprime}a\ expr{\isachardoublequoteclose}%
}
\DefineSnippet{agraph}{
\isacommand{type{\isacharunderscore}synonym}\isamarkupfalse%
\ {\isacharprime}a\ agraph\ {\isacharequal}\ \ {\isachardoublequoteopen}{\isacharparenleft}nat\ {\isasymtimes}\ {\isacharprime}a\ edge{\isacharparenright}\ list{\isachardoublequoteclose}%
}
\DefineSnippet{func}{
\isacommand{datatype}\isamarkupfalse%
\ {\isacharprime}a\ func\ {\isacharequal}\isanewline
\ \ FieldValidFunc\ {\isacharprime}a\ {\isachardoublequoteopen}{\isacharprime}a\ expr{\isachardoublequoteclose}\ {\isacharbar}\isanewline
\ \ FieldAccessFunc\ {\isacharprime}a\ {\isachardoublequoteopen}{\isacharprime}a\ expr{\isachardoublequoteclose}\ {\isacharbar}\isanewline
\ \ VariantValidFunc\ {\isachardoublequoteopen}nat\ list{\isachardoublequoteclose}\ {\isachardoublequoteopen}{\isacharprime}a\ expr{\isachardoublequoteclose}\ {\isacharbar}\isanewline
\ \ VariantAccessFunc\ {\isachardoublequoteopen}nat\ list{\isachardoublequoteclose}\ {\isachardoublequoteopen}{\isacharprime}a\ expr{\isachardoublequoteclose}%
}
\DefineSnippet{get_source}{
\isacommand{fun}\isamarkupfalse%
\ get{\isacharunderscore}source\ {\isacharcolon}{\isacharcolon}\ {\isachardoublequoteopen}{\isacharprime}a\ edge\ {\isasymRightarrow}\ {\isacharprime}a{\isachardoublequoteclose}\ \isakeyword{where}\isanewline
{\isachardoublequoteopen}get{\isacharunderscore}source\ {\isacharparenleft}Edge\ s\ {\isacharunderscore}\ {\isacharunderscore}\ {\isacharunderscore}\ {\isacharunderscore}{\isacharparenright}\ {\isacharequal}\ s{\isachardoublequoteclose}%
}
\DefineSnippet{get_target}{
\isacommand{fun}\isamarkupfalse%
\ get{\isacharunderscore}target\ {\isacharcolon}{\isacharcolon}\ {\isachardoublequoteopen}{\isacharprime}a\ edge\ {\isasymRightarrow}\ {\isacharprime}a{\isachardoublequoteclose}\ \isakeyword{where}\isanewline
{\isachardoublequoteopen}get{\isacharunderscore}target\ {\isacharparenleft}Edge\ {\isacharunderscore}\ t\ {\isacharunderscore}\ {\isacharunderscore}\ {\isacharunderscore}{\isacharparenright}\ {\isacharequal}\ t{\isachardoublequoteclose}%
}
\DefineSnippet{get_condition}{
\isacommand{fun}\isamarkupfalse%
\ get{\isacharunderscore}condition\ {\isacharcolon}{\isacharcolon}\ {\isachardoublequoteopen}{\isacharprime}a\ edge\ {\isasymRightarrow}\ {\isacharprime}a\ expr{\isachardoublequoteclose}\ \isakeyword{where}\isanewline
{\isachardoublequoteopen}get{\isacharunderscore}condition\ {\isacharparenleft}Edge\ {\isacharunderscore}\ {\isacharunderscore}\ c\ {\isacharunderscore}\ {\isacharunderscore}{\isacharparenright}\ {\isacharequal}\ c{\isachardoublequoteclose}%
}
\DefineSnippet{get_length}{
\isacommand{fun}\isamarkupfalse%
\ get{\isacharunderscore}length\ {\isacharcolon}{\isacharcolon}\ {\isachardoublequoteopen}{\isacharprime}a\ edge\ {\isasymRightarrow}\ {\isacharprime}a\ expr{\isachardoublequoteclose}\ \isakeyword{where}\isanewline
{\isachardoublequoteopen}get{\isacharunderscore}length\ {\isacharparenleft}Edge\ {\isacharunderscore}\ {\isacharunderscore}\ {\isacharunderscore}\ l\ {\isacharunderscore}{\isacharparenright}\ {\isacharequal}\ l{\isachardoublequoteclose}%
}
\DefineSnippet{get_first}{
\isacommand{fun}\isamarkupfalse%
\ get{\isacharunderscore}first\ {\isacharcolon}{\isacharcolon}\ {\isachardoublequoteopen}{\isacharprime}a\ edge\ {\isasymRightarrow}\ {\isacharprime}a\ expr{\isachardoublequoteclose}\ \isakeyword{where}\isanewline
{\isachardoublequoteopen}get{\isacharunderscore}first\ {\isacharparenleft}Edge\ {\isacharunderscore}\ {\isacharunderscore}\ {\isacharunderscore}\ {\isacharunderscore}\ f{\isacharparenright}\ {\isacharequal}\ f{\isachardoublequoteclose}%
}
\DefineSnippet{path_edges}{
\isacommand{fun}\isamarkupfalse%
\ path{\isacharunderscore}edges\ {\isacharcolon}{\isacharcolon}\ {\isachardoublequoteopen}{\isacharprime}a\ agraph\ {\isasymRightarrow}\ nat\ list\ {\isasymRightarrow}\ {\isacharprime}a\ edge\ list{\isachardoublequoteclose}\ \isakeyword{where}\isanewline
{\isachardoublequoteopen}path{\isacharunderscore}edges\ {\isacharbrackleft}{\isacharbrackright}\ {\isacharunderscore}\ {\isacharequal}\ {\isacharbrackleft}{\isacharbrackright}{\isachardoublequoteclose}\ {\isacharbar}\isanewline
{\isachardoublequoteopen}path{\isacharunderscore}edges\ {\isacharunderscore}\ {\isacharbrackleft}{\isacharbrackright}\ {\isacharequal}\ {\isacharbrackleft}{\isacharbrackright}{\isachardoublequoteclose}\ {\isacharbar}\isanewline
{\isachardoublequoteopen}path{\isacharunderscore}edges\ g\ {\isacharparenleft}n\ {\isacharhash}\ ns{\isacharparenright}\ {\isacharequal}\ snd\ {\isacharparenleft}g\ {\isacharbang}\ n{\isacharparenright}\ {\isacharhash}\ path{\isacharunderscore}edges\ g\ ns{\isachardoublequoteclose}%
}
\DefineSnippet{outgoing}{
\isacommand{fun}\isamarkupfalse%
\ outgoing\ {\isacharcolon}{\isacharcolon}\ {\isachardoublequoteopen}{\isacharprime}a\ agraph\ {\isasymRightarrow}\ {\isacharprime}a\ {\isasymRightarrow}\ nat\ list{\isachardoublequoteclose}\ \isakeyword{where}\isanewline
{\isachardoublequoteopen}outgoing\ {\isacharbrackleft}{\isacharbrackright}\ n\ {\isacharequal}\ {\isacharbrackleft}{\isacharbrackright}{\isachardoublequoteclose}\ {\isacharbar}\isanewline
{\isachardoublequoteopen}outgoing\ {\isacharparenleft}x\ {\isacharhash}\ xs{\isacharparenright}\ n\ {\isacharequal}\ {\isacharparenleft}if\ n\ {\isacharequal}\ get{\isacharunderscore}source\ {\isacharparenleft}snd\ x{\isacharparenright}\ then\ fst\ x\ {\isacharhash}\ outgoing\ xs\ n\ else\ outgoing\ xs\ n{\isacharparenright}{\isachardoublequoteclose}%
}
\DefineSnippet{incoming}{
\isacommand{fun}\isamarkupfalse%
\ incoming\ {\isacharcolon}{\isacharcolon}\ {\isachardoublequoteopen}{\isacharprime}a\ agraph\ {\isasymRightarrow}\ {\isacharprime}a\ {\isasymRightarrow}\ nat\ list{\isachardoublequoteclose}\ \isakeyword{where}\isanewline
{\isachardoublequoteopen}incoming\ {\isacharbrackleft}{\isacharbrackright}\ n\ {\isacharequal}\ {\isacharbrackleft}{\isacharbrackright}{\isachardoublequoteclose}\ {\isacharbar}\isanewline
{\isachardoublequoteopen}incoming\ {\isacharparenleft}x\ {\isacharhash}\ xs{\isacharparenright}\ n\ {\isacharequal}\ {\isacharparenleft}if\ n\ {\isacharequal}\ get{\isacharunderscore}target\ {\isacharparenleft}snd\ x{\isacharparenright}\ then\ fst\ x\ {\isacharhash}\ incoming\ xs\ n\ else\ incoming\ xs\ n{\isacharparenright}{\isachardoublequoteclose}%
}
\DefineSnippet{revpaths}{
\isacommand{fun}\isamarkupfalse%
\ revpaths\ {\isacharcolon}{\isacharcolon}\ {\isachardoublequoteopen}nat\ {\isasymRightarrow}\ {\isacharprime}a\ agraph\ {\isasymRightarrow}\ nat\ {\isasymRightarrow}\ nat\ list\ list{\isachardoublequoteclose}\ \isakeyword{where}\isanewline
{\isachardoublequoteopen}revpaths\ {\isadigit{0}}\ {\isacharunderscore}\ {\isacharunderscore}\ {\isacharequal}\ {\isacharbrackleft}{\isacharbrackright}{\isachardoublequoteclose}\ {\isacharbar}\isanewline
{\isachardoublequoteopen}revpaths\ {\isacharunderscore}\ {\isacharbrackleft}{\isacharbrackright}\ {\isacharunderscore}\ {\isacharequal}\ {\isacharbrackleft}{\isacharbrackright}{\isachardoublequoteclose}\ {\isacharbar}\isanewline
{\isachardoublequoteopen}revpaths\ i\ g\ e\ {\isacharequal}\ {\isacharparenleft}if\ incoming\ g\ {\isacharparenleft}get{\isacharunderscore}source\ {\isacharparenleft}snd\ {\isacharparenleft}g\ {\isacharbang}\ e{\isacharparenright}{\isacharparenright}{\isacharparenright}\ {\isacharequal}\ {\isacharbrackleft}{\isacharbrackright}\ then\ {\isacharbrackleft}{\isacharbrackleft}e{\isacharbrackright}{\isacharbrackright}\ else\ map\ {\isacharparenleft}{\isacharparenleft}{\isacharhash}{\isacharparenright}\ e{\isacharparenright}\ {\isacharparenleft}concat\ {\isacharparenleft}map\ {\isacharparenleft}revpaths\ {\isacharparenleft}i{\isacharminus}{\isadigit{1}}{\isacharparenright}\ g{\isacharparenright}\ {\isacharparenleft}incoming\ g\ {\isacharparenleft}get{\isacharunderscore}source\ {\isacharparenleft}snd\ {\isacharparenleft}g\ {\isacharbang}\ e{\isacharparenright}{\isacharparenright}{\isacharparenright}{\isacharparenright}{\isacharparenright}{\isacharparenright}{\isacharparenright}{\isachardoublequoteclose}%
}
\DefineSnippet{paths}{
\isacommand{definition}\isamarkupfalse%
\ paths\ {\isacharcolon}{\isacharcolon}\ {\isachardoublequoteopen}{\isacharprime}a\ agraph\ {\isasymRightarrow}\ nat\ {\isasymRightarrow}\ nat\ list\ list{\isachardoublequoteclose}\ \isakeyword{where}\isanewline
{\isachardoublequoteopen}paths\ graph\ i\ {\isacharequal}\ map\ rev\ {\isacharparenleft}revpaths\ {\isadigit{4}}{\isadigit{2}}\ graph\ i{\isacharparenright}{\isachardoublequoteclose}%
}
\DefineSnippet{resolve_length}{
\isacommand{fun}\isamarkupfalse%
\ resolve{\isacharunderscore}length\ {\isacharcolon}{\isacharcolon}\ {\isachardoublequoteopen}{\isacharprime}a\ edge\ list\ {\isasymRightarrow}\ {\isacharprime}a\ {\isasymRightarrow}\ {\isacharprime}a\ expr{\isachardoublequoteclose}\ \isakeyword{where}\isanewline
{\isachardoublequoteopen}resolve{\isacharunderscore}length\ {\isacharbrackleft}{\isacharbrackright}\ {\isacharunderscore}\ {\isacharequal}\ undefined{\isachardoublequoteclose}\ {\isacharbar}\isanewline
{\isachardoublequoteopen}resolve{\isacharunderscore}length\ {\isacharparenleft}x\ {\isacharhash}\ xs{\isacharparenright}\ v\ {\isacharequal}\ {\isacharparenleft}if\ get{\isacharunderscore}target\ x\ {\isacharequal}\ v\ then\ get{\isacharunderscore}length\ x\ else\ resolve{\isacharunderscore}length\ xs\ v{\isacharparenright}{\isachardoublequoteclose}%
}
\DefineSnippet{resolve_first}{
\isacommand{fun}\isamarkupfalse%
\ resolve{\isacharunderscore}first\ {\isacharcolon}{\isacharcolon}\ {\isachardoublequoteopen}{\isacharprime}a\ edge\ list\ {\isasymRightarrow}\ {\isacharprime}a\ {\isasymRightarrow}\ {\isacharprime}a\ expr{\isachardoublequoteclose}\ \isakeyword{where}\isanewline
{\isachardoublequoteopen}resolve{\isacharunderscore}first\ {\isacharbrackleft}{\isacharbrackright}\ {\isacharunderscore}\ {\isacharequal}\ undefined{\isachardoublequoteclose}\ {\isacharbar}\isanewline
{\isachardoublequoteopen}resolve{\isacharunderscore}first\ {\isacharparenleft}x\ {\isacharhash}\ xs{\isacharparenright}\ v\ {\isacharequal}\ {\isacharparenleft}if\ get{\isacharunderscore}target\ x\ {\isacharequal}\ v\ then\ get{\isacharunderscore}first\ x\ else\ resolve{\isacharunderscore}first\ xs\ v{\isacharparenright}{\isachardoublequoteclose}%
}
\DefineSnippet{msubs}{
\isacommand{fun}\isamarkupfalse%
\ msubs\ {\isacharcolon}{\isacharcolon}\ {\isachardoublequoteopen}nat\ {\isasymRightarrow}\ {\isacharprime}a\ edge\ list\ {\isasymRightarrow}\ {\isacharprime}a\ expr\ {\isasymRightarrow}\ {\isacharprime}a\ expr{\isachardoublequoteclose}\ \isakeyword{where}\isanewline
{\isachardoublequoteopen}msubs\ {\isadigit{0}}\ {\isacharunderscore}\ {\isacharunderscore}\ {\isacharequal}\ undefined{\isachardoublequoteclose}\ {\isacharbar}\isanewline
{\isachardoublequoteopen}msubs\ i\ p\ {\isacharparenleft}Add\ lhs\ rhs{\isacharparenright}\ {\isacharequal}\ {\isacharparenleft}Add\ {\isacharparenleft}msubs\ {\isacharparenleft}i\ {\isacharminus}\ {\isadigit{1}}{\isacharparenright}\ p\ lhs{\isacharparenright}\ {\isacharparenleft}msubs\ {\isacharparenleft}i\ {\isacharminus}\ {\isadigit{1}}{\isacharparenright}\ p\ rhs{\isacharparenright}{\isacharparenright}{\isachardoublequoteclose}\ {\isacharbar}\isanewline
{\isachardoublequoteopen}msubs\ i\ p\ {\isacharparenleft}Mul\ lhs\ rhs{\isacharparenright}\ {\isacharequal}\ {\isacharparenleft}Mul\ {\isacharparenleft}msubs\ {\isacharparenleft}i\ {\isacharminus}\ {\isadigit{1}}{\isacharparenright}\ p\ lhs{\isacharparenright}\ {\isacharparenleft}msubs\ {\isacharparenleft}i\ {\isacharminus}\ {\isadigit{1}}{\isacharparenright}\ p\ rhs{\isacharparenright}{\isacharparenright}{\isachardoublequoteclose}\ {\isacharbar}\isanewline
{\isachardoublequoteopen}msubs\ i\ p\ {\isacharparenleft}Sub\ lhs\ rhs{\isacharparenright}\ {\isacharequal}\ {\isacharparenleft}Sub\ {\isacharparenleft}msubs\ {\isacharparenleft}i\ {\isacharminus}\ {\isadigit{1}}{\isacharparenright}\ p\ lhs{\isacharparenright}\ {\isacharparenleft}msubs\ {\isacharparenleft}i\ {\isacharminus}\ {\isadigit{1}}{\isacharparenright}\ p\ rhs{\isacharparenright}{\isacharparenright}{\isachardoublequoteclose}\ {\isacharbar}\isanewline
{\isachardoublequoteopen}msubs\ i\ p\ {\isacharparenleft}Value\ lhs\ rhs{\isacharparenright}\ {\isacharequal}\ Value\ {\isacharparenleft}msubs\ {\isacharparenleft}i\ {\isacharminus}\ {\isadigit{1}}{\isacharparenright}\ p\ lhs{\isacharparenright}\ {\isacharparenleft}msubs\ {\isacharparenleft}i\ {\isacharminus}\ {\isadigit{1}}{\isacharparenright}\ p\ rhs{\isacharparenright}{\isachardoublequoteclose}\ {\isacharbar}\isanewline
{\isachardoublequoteopen}msubs\ i\ p\ {\isacharparenleft}FieldValue\ l{\isacharparenright}\ {\isacharequal}\ msubs\ {\isacharparenleft}i\ {\isacharminus}\ {\isadigit{1}}{\isacharparenright}\ p\ {\isacharparenleft}Value\ {\isacharparenleft}FieldFirst\ l{\isacharparenright}\ {\isacharparenleft}FieldLength\ l{\isacharparenright}{\isacharparenright}{\isachardoublequoteclose}\ {\isacharbar}\isanewline
{\isachardoublequoteopen}msubs\ i\ p\ {\isacharparenleft}FieldLength\ l{\isacharparenright}\ {\isacharequal}\ msubs\ {\isacharparenleft}i\ {\isacharminus}\ {\isadigit{1}}{\isacharparenright}\ p\ {\isacharparenleft}resolve{\isacharunderscore}length\ p\ l{\isacharparenright}{\isachardoublequoteclose}\ {\isacharbar}\isanewline
{\isachardoublequoteopen}msubs\ i\ p\ {\isacharparenleft}FieldFirst\ l{\isacharparenright}\ {\isacharequal}\ msubs\ {\isacharparenleft}i\ {\isacharminus}\ {\isadigit{1}}{\isacharparenright}\ p\ {\isacharparenleft}resolve{\isacharunderscore}first\ p\ l{\isacharparenright}{\isachardoublequoteclose}\ {\isacharbar}\isanewline
{\isachardoublequoteopen}msubs\ i\ p\ e\ {\isacharequal}\ e{\isachardoublequoteclose}%
}
\DefineSnippet{bsubs}{
\isacommand{fun}\isamarkupfalse%
\ bsubs\ {\isacharcolon}{\isacharcolon}\ {\isachardoublequoteopen}nat\ {\isasymRightarrow}\ {\isacharprime}a\ edge\ list\ {\isasymRightarrow}\ {\isacharprime}a\ expr\ {\isasymRightarrow}\ {\isacharprime}a\ expr{\isachardoublequoteclose}\ \isakeyword{where}\isanewline
{\isachardoublequoteopen}bsubs\ {\isadigit{0}}\ {\isacharunderscore}\ {\isacharunderscore}\ {\isacharequal}\ undefined{\isachardoublequoteclose}\ {\isacharbar}\isanewline
{\isachardoublequoteopen}bsubs\ i\ p\ {\isacharparenleft}And\ lhs\ rhs{\isacharparenright}\ {\isacharequal}\ And\ {\isacharparenleft}bsubs\ {\isacharparenleft}i\ {\isacharminus}\ {\isadigit{1}}{\isacharparenright}\ p\ lhs{\isacharparenright}\ {\isacharparenleft}bsubs\ {\isacharparenleft}i\ {\isacharminus}\ {\isadigit{1}}{\isacharparenright}\ p\ rhs{\isacharparenright}{\isachardoublequoteclose}\ {\isacharbar}\isanewline
{\isachardoublequoteopen}bsubs\ i\ p\ {\isacharparenleft}Or\ lhs\ rhs{\isacharparenright}\ {\isacharequal}\ Or\ {\isacharparenleft}bsubs\ {\isacharparenleft}i\ {\isacharminus}\ {\isadigit{1}}{\isacharparenright}\ p\ lhs{\isacharparenright}\ {\isacharparenleft}bsubs\ {\isacharparenleft}i\ {\isacharminus}\ {\isadigit{1}}{\isacharparenright}\ p\ rhs{\isacharparenright}{\isachardoublequoteclose}\ {\isacharbar}\isanewline
{\isachardoublequoteopen}bsubs\ i\ p\ {\isacharparenleft}Eq\ lhs\ rhs{\isacharparenright}\ {\isacharequal}\ Eq\ {\isacharparenleft}msubs\ {\isacharparenleft}i\ {\isacharminus}\ {\isadigit{1}}{\isacharparenright}\ p\ lhs{\isacharparenright}\ {\isacharparenleft}msubs\ {\isacharparenleft}i\ {\isacharminus}\ {\isadigit{1}}{\isacharparenright}\ p\ rhs{\isacharparenright}{\isachardoublequoteclose}\ {\isacharbar}\isanewline
{\isachardoublequoteopen}bsubs\ i\ p\ {\isacharparenleft}Ne\ lhs\ rhs{\isacharparenright}\ {\isacharequal}\ Ne\ {\isacharparenleft}msubs\ {\isacharparenleft}i\ {\isacharminus}\ {\isadigit{1}}{\isacharparenright}\ p\ lhs{\isacharparenright}\ {\isacharparenleft}msubs\ {\isacharparenleft}i\ {\isacharminus}\ {\isadigit{1}}{\isacharparenright}\ p\ rhs{\isacharparenright}{\isachardoublequoteclose}\ {\isacharbar}\isanewline
{\isachardoublequoteopen}bsubs\ i\ p\ {\isacharparenleft}Lt\ lhs\ rhs{\isacharparenright}\ {\isacharequal}\ Lt\ {\isacharparenleft}msubs\ {\isacharparenleft}i\ {\isacharminus}\ {\isadigit{1}}{\isacharparenright}\ p\ lhs{\isacharparenright}\ {\isacharparenleft}msubs\ {\isacharparenleft}i\ {\isacharminus}\ {\isadigit{1}}{\isacharparenright}\ p\ rhs{\isacharparenright}{\isachardoublequoteclose}\ {\isacharbar}\isanewline
{\isachardoublequoteopen}bsubs\ i\ p\ {\isacharparenleft}Le\ lhs\ rhs{\isacharparenright}\ {\isacharequal}\ Le\ {\isacharparenleft}msubs\ {\isacharparenleft}i\ {\isacharminus}\ {\isadigit{1}}{\isacharparenright}\ p\ lhs{\isacharparenright}\ {\isacharparenleft}msubs\ {\isacharparenleft}i\ {\isacharminus}\ {\isadigit{1}}{\isacharparenright}\ p\ rhs{\isacharparenright}{\isachardoublequoteclose}\ {\isacharbar}\isanewline
{\isachardoublequoteopen}bsubs\ i\ p\ {\isacharparenleft}Gt\ lhs\ rhs{\isacharparenright}\ {\isacharequal}\ Gt\ {\isacharparenleft}msubs\ {\isacharparenleft}i\ {\isacharminus}\ {\isadigit{1}}{\isacharparenright}\ p\ lhs{\isacharparenright}\ {\isacharparenleft}msubs\ {\isacharparenleft}i\ {\isacharminus}\ {\isadigit{1}}{\isacharparenright}\ p\ rhs{\isacharparenright}{\isachardoublequoteclose}\ {\isacharbar}\isanewline
{\isachardoublequoteopen}bsubs\ i\ p\ {\isacharparenleft}Ge\ lhs\ rhs{\isacharparenright}\ {\isacharequal}\ Ge\ {\isacharparenleft}msubs\ {\isacharparenleft}i\ {\isacharminus}\ {\isadigit{1}}{\isacharparenright}\ p\ lhs{\isacharparenright}\ {\isacharparenleft}msubs\ {\isacharparenleft}i\ {\isacharminus}\ {\isadigit{1}}{\isacharparenright}\ p\ rhs{\isacharparenright}{\isachardoublequoteclose}\ {\isacharbar}\isanewline
{\isachardoublequoteopen}bsubs\ i\ p\ e\ {\isacharequal}\ e{\isachardoublequoteclose}%
}
\DefineSnippet{subs}{
\isacommand{definition}\isamarkupfalse%
\ subs\ {\isacharcolon}{\isacharcolon}\ {\isachardoublequoteopen}{\isacharprime}a\ edge\ list\ {\isasymRightarrow}\ {\isacharprime}a\ expr\ {\isasymRightarrow}\ {\isacharprime}a\ expr{\isachardoublequoteclose}\ \isakeyword{where}\isanewline
{\isachardoublequoteopen}subs\ p\ e\ {\isacharequal}\ msubs\ {\isadigit{4}}{\isadigit{2}}\ p\ {\isacharparenleft}bsubs\ {\isadigit{4}}{\isadigit{2}}\ p\ e{\isacharparenright}{\isachardoublequoteclose}%
}
\DefineSnippet{simplify_math'}{
\isacommand{fun}\isamarkupfalse%
\ simplify{\isacharunderscore}math{\isacharunderscore}add\ {\isacharcolon}{\isacharcolon}\ {\isachardoublequoteopen}{\isacharprime}a\ expr\ {\isasymRightarrow}\ {\isacharprime}a\ expr{\isachardoublequoteclose}\ \isakeyword{where}\isanewline
{\isachardoublequoteopen}simplify{\isacharunderscore}math{\isacharunderscore}add\ {\isacharparenleft}Add\ {\isacharparenleft}Num\ a{\isacharparenright}\ {\isacharparenleft}Num\ b{\isacharparenright}{\isacharparenright}\ {\isacharequal}\ Num\ {\isacharparenleft}a\ {\isacharplus}\ b{\isacharparenright}{\isachardoublequoteclose}\ {\isacharbar}\isanewline
{\isachardoublequoteopen}simplify{\isacharunderscore}math{\isacharunderscore}add\ e\ {\isacharequal}\ e{\isachardoublequoteclose}\isanewline
\isanewline
\isacommand{fun}\isamarkupfalse%
\ simplify{\isacharunderscore}math{\isacharunderscore}mul\ {\isacharcolon}{\isacharcolon}\ {\isachardoublequoteopen}{\isacharprime}a\ expr\ {\isasymRightarrow}\ {\isacharprime}a\ expr{\isachardoublequoteclose}\ \isakeyword{where}\isanewline
{\isachardoublequoteopen}simplify{\isacharunderscore}math{\isacharunderscore}mul\ {\isacharparenleft}Mul\ {\isacharparenleft}Num\ a{\isacharparenright}\ {\isacharparenleft}Num\ b{\isacharparenright}{\isacharparenright}\ {\isacharequal}\ Num\ {\isacharparenleft}a\ {\isacharasterisk}\ b{\isacharparenright}{\isachardoublequoteclose}\ {\isacharbar}\isanewline
{\isachardoublequoteopen}simplify{\isacharunderscore}math{\isacharunderscore}mul\ e\ {\isacharequal}\ e{\isachardoublequoteclose}\isanewline
\isanewline
\isacommand{fun}\isamarkupfalse%
\ simplify{\isacharunderscore}math{\isacharunderscore}sub\ {\isacharcolon}{\isacharcolon}\ {\isachardoublequoteopen}{\isacharprime}a\ expr\ {\isasymRightarrow}\ {\isacharprime}a\ expr{\isachardoublequoteclose}\ \isakeyword{where}\isanewline
{\isachardoublequoteopen}simplify{\isacharunderscore}math{\isacharunderscore}sub\ {\isacharparenleft}Sub\ {\isacharparenleft}Num\ a{\isacharparenright}\ {\isacharparenleft}Num\ b{\isacharparenright}{\isacharparenright}\ {\isacharequal}\ Num\ {\isacharparenleft}a\ {\isacharminus}\ b{\isacharparenright}{\isachardoublequoteclose}\ {\isacharbar}\isanewline
{\isachardoublequoteopen}simplify{\isacharunderscore}math{\isacharunderscore}sub\ e\ {\isacharequal}\ e{\isachardoublequoteclose}%
}
\DefineSnippet{simplify_bool'}{
\isacommand{fun}\isamarkupfalse%
\ simplify{\isacharunderscore}bool{\isacharprime}\ {\isacharcolon}{\isacharcolon}\ {\isachardoublequoteopen}{\isacharprime}a\ expr\ {\isasymRightarrow}\ {\isacharprime}a\ expr{\isachardoublequoteclose}\ \isakeyword{where}\isanewline
{\isachardoublequoteopen}simplify{\isacharunderscore}bool{\isacharprime}\ {\isacharparenleft}And\ a\ b{\isacharparenright}\ {\isacharequal}\ {\isacharparenleft}case\ b\ of\ True\ {\isacharequal}{\isachargreater}\ a\ {\isacharbar}\ {\isacharunderscore}\ {\isacharequal}{\isachargreater}\ {\isacharparenleft}case\ a\ of\ True\ {\isacharequal}{\isachargreater}\ b\ {\isacharbar}\ {\isacharunderscore}\ {\isacharequal}{\isachargreater}\ And\ a\ b{\isacharparenright}{\isacharparenright}{\isachardoublequoteclose}\ {\isacharbar}\isanewline
{\isachardoublequoteopen}simplify{\isacharunderscore}bool{\isacharprime}\ {\isacharparenleft}Or\ a\ b{\isacharparenright}\ {\isacharequal}\ {\isacharparenleft}case\ b\ of\ True\ {\isacharequal}{\isachargreater}\ True\ {\isacharbar}\ {\isacharunderscore}\ {\isacharequal}{\isachargreater}\ {\isacharparenleft}case\ a\ of\ True\ {\isacharequal}{\isachargreater}\ True\ {\isacharbar}\ {\isacharunderscore}\ {\isacharequal}{\isachargreater}\ Or\ a\ b{\isacharparenright}{\isacharparenright}{\isachardoublequoteclose}\ {\isacharbar}\isanewline
{\isachardoublequoteopen}simplify{\isacharunderscore}bool{\isacharprime}\ e\ {\isacharequal}\ e{\isachardoublequoteclose}%
}
\DefineSnippet{simplify_math}{
\isacommand{fun}\isamarkupfalse%
\ simplify{\isacharunderscore}math\ {\isacharcolon}{\isacharcolon}\ {\isachardoublequoteopen}{\isacharprime}a\ expr\ {\isasymRightarrow}\ {\isacharprime}a\ expr{\isachardoublequoteclose}\ \isakeyword{where}\isanewline
{\isachardoublequoteopen}simplify{\isacharunderscore}math\ {\isacharparenleft}Add\ a\ b{\isacharparenright}\ {\isacharequal}\ simplify{\isacharunderscore}math{\isacharunderscore}add\ {\isacharparenleft}Add\ {\isacharparenleft}simplify{\isacharunderscore}math\ a{\isacharparenright}\ {\isacharparenleft}simplify{\isacharunderscore}math\ \ b{\isacharparenright}{\isacharparenright}{\isachardoublequoteclose}\ {\isacharbar}\isanewline
{\isachardoublequoteopen}simplify{\isacharunderscore}math\ {\isacharparenleft}Mul\ a\ b{\isacharparenright}\ {\isacharequal}\ simplify{\isacharunderscore}math{\isacharunderscore}mul\ {\isacharparenleft}Mul\ {\isacharparenleft}simplify{\isacharunderscore}math\ a{\isacharparenright}\ {\isacharparenleft}simplify{\isacharunderscore}math\ \ b{\isacharparenright}{\isacharparenright}{\isachardoublequoteclose}\ {\isacharbar}\isanewline
{\isachardoublequoteopen}simplify{\isacharunderscore}math\ {\isacharparenleft}Sub\ a\ b{\isacharparenright}\ {\isacharequal}\ simplify{\isacharunderscore}math{\isacharunderscore}sub\ {\isacharparenleft}Sub\ {\isacharparenleft}simplify{\isacharunderscore}math\ a{\isacharparenright}\ {\isacharparenleft}simplify{\isacharunderscore}math\ \ b{\isacharparenright}{\isacharparenright}{\isachardoublequoteclose}\ {\isacharbar}\isanewline
{\isachardoublequoteopen}simplify{\isacharunderscore}math\ {\isacharparenleft}Value\ a\ b{\isacharparenright}\ {\isacharequal}\ Value\ {\isacharparenleft}simplify{\isacharunderscore}math\ a{\isacharparenright}\ {\isacharparenleft}simplify{\isacharunderscore}math\ \ b{\isacharparenright}{\isachardoublequoteclose}\ {\isacharbar}\isanewline
{\isachardoublequoteopen}simplify{\isacharunderscore}math\ e\ {\isacharequal}\ e{\isachardoublequoteclose}%
}
\DefineSnippet{simplify_bool}{
\isacommand{fun}\isamarkupfalse%
\ simplify{\isacharunderscore}bool\ {\isacharcolon}{\isacharcolon}\ {\isachardoublequoteopen}{\isacharprime}a\ expr\ {\isasymRightarrow}\ {\isacharprime}a\ expr{\isachardoublequoteclose}\ \isakeyword{where}\isanewline
{\isachardoublequoteopen}simplify{\isacharunderscore}bool\ {\isacharparenleft}And\ a\ b{\isacharparenright}\ {\isacharequal}\ simplify{\isacharunderscore}bool{\isacharprime}\ {\isacharparenleft}And\ {\isacharparenleft}simplify{\isacharunderscore}bool\ a{\isacharparenright}\ {\isacharparenleft}simplify{\isacharunderscore}bool\ \ b{\isacharparenright}{\isacharparenright}{\isachardoublequoteclose}\ {\isacharbar}\isanewline
{\isachardoublequoteopen}simplify{\isacharunderscore}bool\ {\isacharparenleft}Or\ a\ b{\isacharparenright}\ {\isacharequal}\ simplify{\isacharunderscore}bool{\isacharprime}\ {\isacharparenleft}Or\ {\isacharparenleft}simplify{\isacharunderscore}bool\ a{\isacharparenright}\ {\isacharparenleft}simplify{\isacharunderscore}bool\ \ b{\isacharparenright}{\isacharparenright}{\isachardoublequoteclose}\ {\isacharbar}\isanewline
{\isachardoublequoteopen}simplify{\isacharunderscore}bool\ {\isacharparenleft}Eq\ a\ b{\isacharparenright}\ {\isacharequal}\ Eq\ {\isacharparenleft}simplify{\isacharunderscore}math\ a{\isacharparenright}\ {\isacharparenleft}simplify{\isacharunderscore}math\ \ b{\isacharparenright}{\isachardoublequoteclose}\ {\isacharbar}\isanewline
{\isachardoublequoteopen}simplify{\isacharunderscore}bool\ {\isacharparenleft}Ne\ a\ b{\isacharparenright}\ {\isacharequal}\ Ne\ {\isacharparenleft}simplify{\isacharunderscore}math\ a{\isacharparenright}\ {\isacharparenleft}simplify{\isacharunderscore}math\ \ b{\isacharparenright}{\isachardoublequoteclose}\ {\isacharbar}\isanewline
{\isachardoublequoteopen}simplify{\isacharunderscore}bool\ {\isacharparenleft}Lt\ a\ b{\isacharparenright}\ {\isacharequal}\ Lt\ {\isacharparenleft}simplify{\isacharunderscore}math\ a{\isacharparenright}\ {\isacharparenleft}simplify{\isacharunderscore}math\ \ b{\isacharparenright}{\isachardoublequoteclose}\ {\isacharbar}\isanewline
{\isachardoublequoteopen}simplify{\isacharunderscore}bool\ {\isacharparenleft}Le\ a\ b{\isacharparenright}\ {\isacharequal}\ Le\ {\isacharparenleft}simplify{\isacharunderscore}math\ a{\isacharparenright}\ {\isacharparenleft}simplify{\isacharunderscore}math\ \ b{\isacharparenright}{\isachardoublequoteclose}\ {\isacharbar}\isanewline
{\isachardoublequoteopen}simplify{\isacharunderscore}bool\ {\isacharparenleft}Gt\ a\ b{\isacharparenright}\ {\isacharequal}\ Gt\ {\isacharparenleft}simplify{\isacharunderscore}math\ a{\isacharparenright}\ {\isacharparenleft}simplify{\isacharunderscore}math\ \ b{\isacharparenright}{\isachardoublequoteclose}\ {\isacharbar}\isanewline
{\isachardoublequoteopen}simplify{\isacharunderscore}bool\ {\isacharparenleft}Ge\ a\ b{\isacharparenright}\ {\isacharequal}\ Ge\ {\isacharparenleft}simplify{\isacharunderscore}math\ a{\isacharparenright}\ {\isacharparenleft}simplify{\isacharunderscore}math\ \ b{\isacharparenright}{\isachardoublequoteclose}\ {\isacharbar}\isanewline
{\isachardoublequoteopen}simplify{\isacharunderscore}bool\ e\ {\isacharequal}\ e{\isachardoublequoteclose}%
}
\DefineSnippet{path_attrs}{
\isacommand{definition}\isamarkupfalse%
\ path{\isacharunderscore}attrs\ {\isacharcolon}{\isacharcolon}\ {\isachardoublequoteopen}{\isacharprime}a\ agraph\ {\isasymRightarrow}\ {\isacharparenleft}nat\ list\ {\isasymtimes}\ {\isacharprime}a\ expr\ {\isasymtimes}\ {\isacharprime}a\ expr\ {\isasymtimes}\ {\isacharprime}a\ expr{\isacharparenright}\ list{\isachardoublequoteclose}\ \isakeyword{where}\isanewline
{\isachardoublequoteopen}path{\isacharunderscore}attrs\ graph\ {\isacharequal}\isanewline
\ \ {\isacharbrackleft}let\ edges\ {\isacharequal}\ path{\isacharunderscore}edges\ graph\ path\ in\isanewline
\ \ \ \ \ {\isacharparenleft}path{\isacharcomma}\ subs\ edges\ {\isacharparenleft}get{\isacharunderscore}condition\ {\isacharparenleft}last\ edges{\isacharparenright}{\isacharparenright}{\isacharcomma}\isanewline
\ \ \ \ \ \ subs\ edges\ {\isacharparenleft}get{\isacharunderscore}length\ {\isacharparenleft}last\ edges{\isacharparenright}{\isacharparenright}{\isacharcomma}\isanewline
\ \ \ \ \ \ subs\ edges\ {\isacharparenleft}get{\isacharunderscore}first\ {\isacharparenleft}last\ edges{\isacharparenright}{\isacharparenright}{\isacharparenright}\isanewline
\ \ \ {\isachardot}\ path\ {\isasymleftarrow}\ concat\ {\isacharbrackleft}paths\ graph\ i\ {\isachardot}\ i\ {\isasymleftarrow}\ map\ fst\ graph{\isacharbrackright}{\isacharbrackright}{\isachardoublequoteclose}%
}
\DefineSnippet{graph_nodes}{
\isacommand{fun}\isamarkupfalse%
\ graph{\isacharunderscore}nodes\ {\isacharcolon}{\isacharcolon}\ {\isachardoublequoteopen}{\isacharprime}a\ agraph\ {\isasymRightarrow}\ {\isacharprime}a\ set\ {\isasymRightarrow}\ {\isacharprime}a\ list{\isachardoublequoteclose}\ \isakeyword{where}\isanewline
{\isachardoublequoteopen}graph{\isacharunderscore}nodes\ {\isacharbrackleft}{\isacharbrackright}\ {\isacharunderscore}\ {\isacharequal}\ {\isacharbrackleft}{\isacharbrackright}{\isachardoublequoteclose}\ {\isacharbar}\isanewline
{\isachardoublequoteopen}graph{\isacharunderscore}nodes\ {\isacharparenleft}x\ {\isacharhash}\ xs{\isacharparenright}\ s\ {\isacharequal}\ {\isacharparenleft}if\ get{\isacharunderscore}target\ {\isacharparenleft}snd\ x{\isacharparenright}\ {\isasymnotin}\ s\ then\ get{\isacharunderscore}target\ {\isacharparenleft}snd\ x{\isacharparenright}\ {\isacharhash}\ graph{\isacharunderscore}nodes\ xs\ {\isacharparenleft}insert\ {\isacharparenleft}get{\isacharunderscore}target\ {\isacharparenleft}snd\ x{\isacharparenright}{\isacharparenright}\ s{\isacharparenright}\ else\ graph{\isacharunderscore}nodes\ xs\ {\isacharparenleft}insert\ {\isacharparenleft}get{\isacharunderscore}target\ {\isacharparenleft}snd\ x{\isacharparenright}{\isacharparenright}\ s{\isacharparenright}{\isacharparenright}\ {\isachardoublequoteclose}%
}
\DefineSnippet{path_conds}{
\isacommand{fun}\isamarkupfalse%
\ path{\isacharunderscore}conds\ {\isacharcolon}{\isacharcolon}\ {\isachardoublequoteopen}{\isacharprime}a\ edge\ list\ {\isasymRightarrow}\ {\isacharprime}a\ expr\ list{\isachardoublequoteclose}\ \isakeyword{where}\isanewline
{\isachardoublequoteopen}path{\isacharunderscore}conds\ {\isacharbrackleft}{\isacharbrackright}\ {\isacharequal}\ {\isacharbrackleft}{\isacharbrackright}{\isachardoublequoteclose}\ {\isacharbar}\isanewline
{\isachardoublequoteopen}path{\isacharunderscore}conds\ {\isacharparenleft}x\ {\isacharhash}\ xs{\isacharparenright}\ {\isacharequal}\ get{\isacharunderscore}condition\ x\ {\isacharhash}\ path{\isacharunderscore}conds\ xs{\isachardoublequoteclose}%
}
\DefineSnippet{any}{
\isacommand{fun}\isamarkupfalse%
\ any\ {\isacharcolon}{\isacharcolon}\ {\isachardoublequoteopen}{\isacharprime}a\ expr\ list\ {\isasymRightarrow}\ {\isacharprime}a\ expr{\isachardoublequoteclose}\ \isakeyword{where}\isanewline
{\isachardoublequoteopen}any\ {\isacharbrackleft}{\isacharbrackright}\ {\isacharequal}\ True{\isachardoublequoteclose}\ {\isacharbar}\isanewline
{\isachardoublequoteopen}any\ {\isacharparenleft}x\ {\isacharhash}\ {\isacharbrackleft}{\isacharbrackright}{\isacharparenright}\ {\isacharequal}\ x{\isachardoublequoteclose}\ {\isacharbar}\isanewline
{\isachardoublequoteopen}any\ {\isacharparenleft}x\ {\isacharhash}\ xs{\isacharparenright}\ {\isacharequal}\ Or\ x\ {\isacharparenleft}any\ xs{\isacharparenright}{\isachardoublequoteclose}%
}
\DefineSnippet{node_paths}{
\isacommand{definition}\isamarkupfalse%
\ node{\isacharunderscore}paths\ {\isacharcolon}{\isacharcolon}\ {\isachardoublequoteopen}{\isacharprime}a\ agraph\ {\isasymRightarrow}\ {\isacharparenleft}{\isacharprime}a\ {\isasymtimes}\ {\isacharparenleft}nat\ list\ {\isasymtimes}\ {\isacharprime}a\ expr{\isacharparenright}\ list{\isacharparenright}\ list{\isachardoublequoteclose}\ \isakeyword{where}\isanewline
{\isachardoublequoteopen}node{\isacharunderscore}paths\ graph\ \ {\isacharequal}\ \isanewline
\ \ {\isacharbrackleft}{\isacharparenleft}node{\isacharcomma}\isanewline
\ \ \ \ concat\ {\isacharbrackleft}{\isacharbrackleft}{\isacharparenleft}path{\isacharcomma}\isanewline
\ \ \ \ \ \ \ \ \ \ \ \ \ \ subs\ {\isacharparenleft}path{\isacharunderscore}edges\ graph\ path{\isacharparenright}\isanewline
\ \ \ \ \ \ \ \ \ \ \ \ \ \ \ \ \ \ \ {\isacharparenleft}any\ {\isacharparenleft}path{\isacharunderscore}conds\ {\isacharparenleft}path{\isacharunderscore}edges\ graph\ {\isacharparenleft}outgoing\ graph\ node{\isacharparenright}{\isacharparenright}{\isacharparenright}{\isacharparenright}{\isacharparenright}\isanewline
\ \ \ \ \ \ \ \ \ \ \ \ \ {\isachardot}\ path\ {\isasymleftarrow}\ paths\ graph\ edge{\isacharbrackright}\isanewline
\ \ \ \ \ \ \ \ \ \ \ \ {\isachardot}\ edge\ {\isasymleftarrow}\ incoming\ graph\ node{\isacharbrackright}{\isacharparenright}\isanewline
\ \ \ {\isachardot}\ node\ {\isasymleftarrow}\ graph{\isacharunderscore}nodes\ graph\ {\isacharbraceleft}{\isacharbraceright}{\isacharbrackright}{\isachardoublequoteclose}%
}
\DefineSnippet{valid_calls}{
\isacommand{fun}\isamarkupfalse%
\ valid{\isacharunderscore}calls\ {\isacharcolon}{\isacharcolon}\ {\isachardoublequoteopen}{\isacharparenleft}nat\ list\ {\isasymtimes}\ {\isacharprime}a\ expr{\isacharparenright}\ list\ {\isasymRightarrow}\ {\isacharprime}a\ expr{\isachardoublequoteclose}\ \isakeyword{where}\isanewline
{\isachardoublequoteopen}valid{\isacharunderscore}calls\ {\isacharbrackleft}{\isacharbrackright}\ {\isacharequal}\ True{\isachardoublequoteclose}\ {\isacharbar}\isanewline
{\isachardoublequoteopen}valid{\isacharunderscore}calls\ {\isacharparenleft}{\isacharparenleft}path{\isacharcomma}\ out{\isacharunderscore}cond{\isacharparenright}\ {\isacharhash}\ {\isacharbrackleft}{\isacharbrackright}{\isacharparenright}\ {\isacharequal}\ And\ {\isacharparenleft}VariantValidCall\ path{\isacharparenright}\ out{\isacharunderscore}cond{\isachardoublequoteclose}\ {\isacharbar}\isanewline
{\isachardoublequoteopen}valid{\isacharunderscore}calls\ {\isacharparenleft}{\isacharparenleft}path{\isacharcomma}\ out{\isacharunderscore}cond{\isacharparenright}\ {\isacharhash}\ xs{\isacharparenright}\ {\isacharequal}\ Or\ {\isacharparenleft}And\ {\isacharparenleft}VariantValidCall\ path{\isacharparenright}\ out{\isacharunderscore}cond{\isacharparenright}\isanewline
\ \ \ \ \ \ \ \ \ \ \ \ \ \ \ \ \ \ \ \ \ \ \ \ \ \ \ \ \ \ \ \ \ \ \ \ \ \ \ \ \ \ {\isacharparenleft}valid{\isacharunderscore}calls\ xs{\isacharparenright}{\isachardoublequoteclose}%
}
\DefineSnippet{field_valid_funcs}{
\isacommand{fun}\isamarkupfalse%
\ field{\isacharunderscore}valid{\isacharunderscore}funcs\ {\isacharcolon}{\isacharcolon}\ {\isachardoublequoteopen}{\isacharparenleft}{\isacharprime}a\ {\isasymtimes}\ {\isacharparenleft}nat\ list\ {\isasymtimes}\ {\isacharprime}a\ expr{\isacharparenright}\ list{\isacharparenright}\ list\ {\isasymRightarrow}\ {\isacharprime}a\ func\ list{\isachardoublequoteclose}\ \isakeyword{where}\isanewline
{\isachardoublequoteopen}field{\isacharunderscore}valid{\isacharunderscore}funcs\ {\isacharbrackleft}{\isacharbrackright}\ {\isacharequal}\ {\isacharbrackleft}{\isacharbrackright}{\isachardoublequoteclose}\ {\isacharbar}\isanewline
{\isachardoublequoteopen}field{\isacharunderscore}valid{\isacharunderscore}funcs\ {\isacharparenleft}{\isacharparenleft}path{\isacharcomma}\ path{\isacharunderscore}cond{\isacharparenright}\ {\isacharhash}\ xs{\isacharparenright}\ {\isacharequal}\isanewline
\ \ FieldValidFunc\ path\ {\isacharparenleft}valid{\isacharunderscore}calls\ path{\isacharunderscore}cond{\isacharparenright}\ {\isacharhash}\ field{\isacharunderscore}valid{\isacharunderscore}funcs\ xs{\isachardoublequoteclose}%
}
\DefineSnippet{access_calls}{
\isacommand{fun}\isamarkupfalse%
\ access{\isacharunderscore}calls\ {\isacharcolon}{\isacharcolon}\ {\isachardoublequoteopen}{\isacharparenleft}nat\ list\ {\isasymtimes}\ {\isacharprime}a\ expr{\isacharparenright}\ list\ {\isasymRightarrow}\ {\isacharprime}a\ expr{\isachardoublequoteclose}\ \isakeyword{where}\isanewline
{\isachardoublequoteopen}access{\isacharunderscore}calls\ {\isacharbrackleft}{\isacharbrackright}\ {\isacharequal}\ Null{\isachardoublequoteclose}\ {\isacharbar}\isanewline
{\isachardoublequoteopen}access{\isacharunderscore}calls\ {\isacharparenleft}{\isacharparenleft}path{\isacharcomma}\ {\isacharunderscore}{\isacharparenright}\ {\isacharhash}\ xs{\isacharparenright}\ {\isacharequal}\isanewline
\ \ IfThenElse\ {\isacharparenleft}VariantValidCall\ path{\isacharparenright}\ {\isacharparenleft}VariantAccessCall\ path{\isacharparenright}\ {\isacharparenleft}access{\isacharunderscore}calls\ xs{\isacharparenright}{\isachardoublequoteclose}%
}
\DefineSnippet{field_access_funcs}{
\isacommand{fun}\isamarkupfalse%
\ field{\isacharunderscore}access{\isacharunderscore}funcs\ {\isacharcolon}{\isacharcolon}\ {\isachardoublequoteopen}{\isacharparenleft}{\isacharprime}a\ {\isasymtimes}\ {\isacharparenleft}nat\ list\ {\isasymtimes}\ {\isacharprime}a\ expr{\isacharparenright}\ list{\isacharparenright}\ list\ {\isasymRightarrow}\ {\isacharprime}a\ func\ list{\isachardoublequoteclose}\ \isakeyword{where}\isanewline
{\isachardoublequoteopen}field{\isacharunderscore}access{\isacharunderscore}funcs\ {\isacharbrackleft}{\isacharbrackright}\ {\isacharequal}\ {\isacharbrackleft}{\isacharbrackright}{\isachardoublequoteclose}\ {\isacharbar}\isanewline
{\isachardoublequoteopen}field{\isacharunderscore}access{\isacharunderscore}funcs\ {\isacharparenleft}{\isacharparenleft}path{\isacharcomma}\ path{\isacharunderscore}cond{\isacharparenright}\ {\isacharhash}\ xs{\isacharparenright}\ {\isacharequal}\isanewline
\ \ FieldAccessFunc\ path\ {\isacharparenleft}access{\isacharunderscore}calls\ path{\isacharunderscore}cond{\isacharparenright}\ {\isacharhash}\ field{\isacharunderscore}access{\isacharunderscore}funcs\ xs{\isachardoublequoteclose}%
}
\DefineSnippet{variant_valid_funcs}{
\isacommand{fun}\isamarkupfalse%
\ variant{\isacharunderscore}valid{\isacharunderscore}funcs\ {\isacharcolon}{\isacharcolon}\ {\isachardoublequoteopen}{\isacharparenleft}nat\ list\ {\isasymtimes}\ {\isacharprime}a\ expr\ {\isasymtimes}\ {\isacharprime}a\ expr\ {\isasymtimes}\ {\isacharprime}a\ expr{\isacharparenright}\ list\ {\isasymRightarrow}\ {\isacharprime}a\ func\ list{\isachardoublequoteclose}\ \isakeyword{where}\isanewline
{\isachardoublequoteopen}variant{\isacharunderscore}valid{\isacharunderscore}funcs\ {\isacharbrackleft}{\isacharbrackright}\ {\isacharequal}\ {\isacharbrackleft}{\isacharbrackright}{\isachardoublequoteclose}\ {\isacharbar}\isanewline
{\isachardoublequoteopen}variant{\isacharunderscore}valid{\isacharunderscore}funcs\ {\isacharparenleft}{\isacharparenleft}path{\isacharcomma}\ cond{\isacharcomma}\ len{\isacharcomma}\ first{\isacharparenright}\ {\isacharhash}\ xs{\isacharparenright}\ {\isacharequal}\isanewline
\ \ VariantValidFunc\ path\ {\isacharparenleft}And\ {\isacharparenleft}if\ init\ path\ {\isasymnoteq}\ {\isacharbrackleft}{\isacharbrackright}\ then\ VariantValidCall\ {\isacharparenleft}init\ path{\isacharparenright}\isanewline
\ \ \ \ \ \ \ \ \ \ \ \ \ \ \ \ \ \ \ \ \ \ \ \ \ \ \ \ \ \ \ \ \ \ \ \ \ \ \ \ \ \ \ \ \ \ \ \ \ else\ True{\isacharparenright}\isanewline
\ \ \ \ \ \ \ \ \ \ \ \ \ \ \ \ \ \ \ \ \ \ \ \ \ \ \ \ \ {\isacharparenleft}And\ {\isacharparenleft}Ge\ BufferLength\ {\isacharparenleft}Add\ first\ len{\isacharparenright}{\isacharparenright}\ cond{\isacharparenright}{\isacharparenright}\isanewline
\ \ {\isacharhash}\ variant{\isacharunderscore}valid{\isacharunderscore}funcs\ xs{\isachardoublequoteclose}%
}
\DefineSnippet{variant_access_funcs}{
\isacommand{fun}\isamarkupfalse%
\ variant{\isacharunderscore}access{\isacharunderscore}funcs\ {\isacharcolon}{\isacharcolon}\ {\isachardoublequoteopen}{\isacharparenleft}nat\ list\ {\isasymtimes}\ {\isacharprime}a\ expr\ {\isasymtimes}\ {\isacharprime}a\ expr\ {\isasymtimes}\ {\isacharprime}a\ expr{\isacharparenright}\ list\ {\isasymRightarrow}\ {\isacharprime}a\ func\ list{\isachardoublequoteclose}\ \isakeyword{where}\isanewline
{\isachardoublequoteopen}variant{\isacharunderscore}access{\isacharunderscore}funcs\ {\isacharbrackleft}{\isacharbrackright}\ {\isacharequal}\ {\isacharbrackleft}{\isacharbrackright}{\isachardoublequoteclose}\ {\isacharbar}\isanewline
{\isachardoublequoteopen}variant{\isacharunderscore}access{\isacharunderscore}funcs\ {\isacharparenleft}{\isacharparenleft}path{\isacharcomma}\ {\isacharunderscore}{\isacharcomma}\ len{\isacharcomma}\ first{\isacharparenright}\ {\isacharhash}\ xs{\isacharparenright}\ {\isacharequal}\isanewline
\ \ VariantAccessFunc\ path\ {\isacharparenleft}Value\ first\ len{\isacharparenright}\ {\isacharhash}\ variant{\isacharunderscore}access{\isacharunderscore}funcs\ xs{\isachardoublequoteclose}%
}
\DefineSnippet{field}{
\isacommand{datatype}\isamarkupfalse%
\ {\isacharprime}a\ field\ {\isacharequal}\ Field\ {\isacharprime}a\ {\isachardoublequoteopen}{\isacharprime}a\ expr{\isachardoublequoteclose}%
}
\DefineSnippet{ethernet_v2}{
\isacommand{datatype}\isamarkupfalse%
\ ethernet{\isacharunderscore}v{\isadigit{2}}\ {\isacharequal}\ Destination\ {\isacharbar}\ Source\ {\isacharbar}\ Type\ {\isacharbar}\ Payload%
}
\DefineSnippet{ethernet_v2_frame}{
\isacommand{definition}\isamarkupfalse%
\ ethernet{\isacharunderscore}v{\isadigit{2}}{\isacharunderscore}frame\ {\isacharcolon}{\isacharcolon}\ {\isachardoublequoteopen}ethernet{\isacharunderscore}v{\isadigit{2}}\ field\ list{\isachardoublequoteclose}\ \isakeyword{where}\isanewline
{\isachardoublequoteopen}ethernet{\isacharunderscore}v{\isadigit{2}}{\isacharunderscore}frame\ {\isacharequal}\ {\isacharbrackleft}Field\ Destination\ {\isacharparenleft}Num\ {\isadigit{4}}{\isadigit{8}}{\isacharparenright}{\isacharcomma}\ Field\ Source\ {\isacharparenleft}Num\ {\isadigit{4}}{\isadigit{8}}{\isacharparenright}{\isacharcomma}\ Field\ Type\ {\isacharparenleft}Num\ {\isadigit{1}}{\isadigit{6}}{\isacharparenright}{\isacharcomma}\ Field\ Payload\ {\isacharparenleft}Sub\ MessageLength\ {\isacharparenleft}Num\ {\isadigit{1}}{\isadigit{1}}{\isadigit{2}}{\isacharparenright}{\isacharparenright}{\isacharbrackright}{\isachardoublequoteclose}%
}
\DefineSnippet{ethernet}{
\isacommand{datatype}\isamarkupfalse%
\ ethernet\ {\isacharequal}\ Destination\ {\isacharbar}\ Source\ {\isacharbar}\ Length\ {\isacharbar}\ Payload%
}
\DefineSnippet{ethernet_frame}{
\isacommand{definition}\isamarkupfalse%
\ ethernet{\isacharunderscore}frame\ {\isacharcolon}{\isacharcolon}\ {\isachardoublequoteopen}ethernet\ field\ list{\isachardoublequoteclose}\ \isakeyword{where}\isanewline
{\isachardoublequoteopen}ethernet{\isacharunderscore}frame\ {\isacharequal}\ {\isacharbrackleft}Field\ Destination\ {\isacharparenleft}Num\ {\isadigit{4}}{\isadigit{8}}{\isacharparenright}{\isacharcomma}\ Field\ Source\ {\isacharparenleft}Num\ {\isadigit{4}}{\isadigit{8}}{\isacharparenright}{\isacharcomma}\ Field\ Length\ {\isacharparenleft}Num\ {\isadigit{1}}{\isadigit{6}}{\isacharparenright}{\isacharcomma}\ Field\ Payload\ {\isacharparenleft}Mul\ {\isacharparenleft}FieldValue\ Length{\isacharparenright}\ {\isacharparenleft}Num\ {\isadigit{8}}{\isacharparenright}{\isacharparenright}{\isacharbrackright}{\isachardoublequoteclose}%
}
\DefineSnippet{ethernet_node}{
\isacommand{datatype}\isamarkupfalse%
\ ethernet{\isacharunderscore}node\ {\isacharequal}\ Init\ {\isacharbar}\ Source\ {\isacharbar}\ Destination\ {\isacharbar}\ Type{\isacharunderscore}Length{\isacharunderscore}TPID\ {\isacharbar}\ Type\ {\isacharbar}\ Payload\ {\isacharbar}\ TPID\ {\isacharbar}\ TCI\ {\isacharbar}\ Final%
}
\DefineSnippet{ethernet_graph}{
\isacommand{definition}\isamarkupfalse%
\ ethernet{\isacharunderscore}graph\ {\isacharcolon}{\isacharcolon}\ {\isachardoublequoteopen}ethernet{\isacharunderscore}node\ edge\ list{\isachardoublequoteclose}\ \isakeyword{where}\isanewline
{\isachardoublequoteopen}ethernet{\isacharunderscore}graph\ {\isacharequal}\ {\isacharbrackleft}\isanewline
\ \ Edge\ Init\ Destination\ True\ {\isacharparenleft}Num\ {\isadigit{4}}{\isadigit{8}}{\isacharparenright}\ {\isacharparenleft}Num\ {\isadigit{0}}{\isacharparenright}{\isacharcomma}\isanewline
\ \ Edge\ Destination\ Source\ True\ {\isacharparenleft}Num\ {\isadigit{4}}{\isadigit{8}}{\isacharparenright}\ {\isacharparenleft}Add\ {\isacharparenleft}FieldFirst\ Destination{\isacharparenright}\ {\isacharparenleft}FieldLength\ Destination{\isacharparenright}{\isacharparenright}{\isacharcomma}\isanewline
\ \ Edge\ Source\ Type{\isacharunderscore}Length{\isacharunderscore}TPID\ True\ {\isacharparenleft}Num\ {\isadigit{1}}{\isadigit{6}}{\isacharparenright}\ {\isacharparenleft}Add\ {\isacharparenleft}FieldFirst\ Source{\isacharparenright}\ {\isacharparenleft}FieldLength\ Source{\isacharparenright}{\isacharparenright}{\isacharcomma}\isanewline
\ \ Edge\ Type{\isacharunderscore}Length{\isacharunderscore}TPID\ Type\ {\isacharparenleft}Ne\ {\isacharparenleft}FieldValue\ Type{\isacharunderscore}Length{\isacharunderscore}TPID{\isacharparenright}\ {\isacharparenleft}Num\ {\isadigit{0}}x{\isadigit{8}}{\isadigit{1}}{\isadigit{0}}{\isadigit{0}}{\isacharparenright}{\isacharparenright}\ {\isacharparenleft}Num\ {\isadigit{1}}{\isadigit{6}}{\isacharparenright}\ {\isacharparenleft}FieldFirst\ Type{\isacharunderscore}Length{\isacharunderscore}TPID{\isacharparenright}{\isacharcomma}\isanewline
\ \ Edge\ Type{\isacharunderscore}Length{\isacharunderscore}TPID\ TPID\ {\isacharparenleft}Eq\ {\isacharparenleft}FieldValue\ Type{\isacharunderscore}Length{\isacharunderscore}TPID{\isacharparenright}\ {\isacharparenleft}Num\ {\isadigit{0}}x{\isadigit{8}}{\isadigit{1}}{\isadigit{0}}{\isadigit{0}}{\isacharparenright}{\isacharparenright}\ {\isacharparenleft}Num\ {\isadigit{1}}{\isadigit{6}}{\isacharparenright}\ {\isacharparenleft}FieldFirst\ Type{\isacharunderscore}Length{\isacharunderscore}TPID{\isacharparenright}{\isacharcomma}\isanewline
\ \ Edge\ TPID\ TCI\ True\ {\isacharparenleft}Num\ {\isadigit{1}}{\isadigit{6}}{\isacharparenright}\ {\isacharparenleft}Add\ {\isacharparenleft}FieldFirst\ TPID{\isacharparenright}\ {\isacharparenleft}FieldLength\ TPID{\isacharparenright}{\isacharparenright}{\isacharcomma}\isanewline
\ \ Edge\ TCI\ Type\ True\ {\isacharparenleft}Num\ {\isadigit{1}}{\isadigit{6}}{\isacharparenright}\ {\isacharparenleft}Add\ {\isacharparenleft}FieldFirst\ TCI{\isacharparenright}\ {\isacharparenleft}FieldLength\ TCI{\isacharparenright}{\isacharparenright}{\isacharcomma}\isanewline
\ \ Edge\ Type\ Payload\ {\isacharparenleft}Le\ {\isacharparenleft}FieldValue\ Type{\isacharparenright}\ {\isacharparenleft}Num\ {\isadigit{1}}{\isadigit{5}}{\isadigit{0}}{\isadigit{0}}{\isacharparenright}{\isacharparenright}\ {\isacharparenleft}Mul\ {\isacharparenleft}FieldValue\ Type{\isacharparenright}\ {\isacharparenleft}Num\ {\isadigit{8}}{\isacharparenright}{\isacharparenright}\ {\isacharparenleft}Add\ {\isacharparenleft}FieldFirst\ Type{\isacharparenright}\ {\isacharparenleft}FieldLength\ Type{\isacharparenright}{\isacharparenright}{\isacharcomma}\isanewline
\ \ Edge\ Type\ Payload\ {\isacharparenleft}Ge\ {\isacharparenleft}FieldValue\ Type{\isacharparenright}\ {\isacharparenleft}Num\ {\isadigit{1}}{\isadigit{5}}{\isadigit{3}}{\isadigit{6}}{\isacharparenright}{\isacharparenright}\ {\isacharparenleft}Add\ {\isacharparenleft}Sub\ MessageLast\ {\isacharparenleft}Add\ {\isacharparenleft}FieldFirst\ Type{\isacharparenright}\ {\isacharparenleft}FieldLength\ Type{\isacharparenright}{\isacharparenright}{\isacharparenright}\ {\isacharparenleft}Num\ {\isadigit{1}}{\isacharparenright}{\isacharparenright}\ {\isacharparenleft}Add\ {\isacharparenleft}FieldFirst\ Type{\isacharparenright}\ {\isacharparenleft}FieldLength\ Type{\isacharparenright}{\isacharparenright}{\isacharcomma}\isanewline
\ \ Edge\ Payload\ Final\ {\isacharparenleft}And\ {\isacharparenleft}Ge\ {\isacharparenleft}FieldLength\ Payload{\isacharparenright}\ {\isacharparenleft}Num\ {\isadigit{3}}{\isadigit{6}}{\isadigit{8}}{\isacharparenright}{\isacharparenright}\ {\isacharparenleft}Le\ {\isacharparenleft}FieldLength\ Payload{\isacharparenright}\ {\isacharparenleft}Num\ {\isadigit{1}}{\isadigit{2}}{\isadigit{0}}{\isadigit{0}}{\isadigit{0}}{\isacharparenright}{\isacharparenright}{\isacharparenright}\ {\isacharparenleft}Num\ {\isadigit{0}}{\isacharparenright}\ {\isacharparenleft}Add\ {\isacharparenleft}FieldFirst\ Payload{\isacharparenright}\ {\isacharparenleft}FieldLength\ Payload{\isacharparenright}{\isacharparenright}\isanewline
{\isacharbrackright}{\isachardoublequoteclose}%
}
\DefineSnippet{ethernet_graph_excerpt}{
\isacommand{definition}\isamarkupfalse%
\ ethernet{\isacharunderscore}graph\ {\isacharcolon}{\isacharcolon}\ {\isachardoublequoteopen}ethernet{\isacharunderscore}node\ edge\ list{\isachardoublequoteclose}\ \isakeyword{where}\isanewline
{\isachardoublequoteopen}ethernet{\isacharunderscore}graph\ {\isacharequal}\ {\isacharbrackleft}\isanewline
\ \ Edge\ Init\ Destination\ True\ {\isacharparenleft}Num\ {\isadigit{4}}{\isadigit{8}}{\isacharparenright}\ {\isacharparenleft}Num\ {\isadigit{0}}{\isacharparenright}{\isacharcomma}\isanewline
\ \ Edge\ Destination\ Source\ True\ {\isacharparenleft}Num\ {\isadigit{4}}{\isadigit{8}}{\isacharparenright}\ {\isacharparenleft}Add\ {\isacharparenleft}FieldFirst\ Destination{\isacharparenright}\ {\isacharparenleft}FieldLength\ Destination{\isacharparenright}{\isacharparenright}{\isacharcomma}\isanewline
\ \ Edge\ Source\ Type{\isacharunderscore}Length{\isacharunderscore}TPID\ True\ {\isacharparenleft}Num\ {\isadigit{1}}{\isadigit{6}}{\isacharparenright}\ {\isacharparenleft}Add\ {\isacharparenleft}FieldFirst\ Source{\isacharparenright}\ {\isacharparenleft}FieldLength\ Source{\isacharparenright}{\isacharparenright}{\isacharcomma}\isanewline
\ \ Edge\ Type{\isacharunderscore}Length{\isacharunderscore}TPID\ Type\ {\isacharparenleft}Ne\ {\isacharparenleft}FieldValue\ Type{\isacharunderscore}Length{\isacharunderscore}TPID{\isacharparenright}\ {\isacharparenleft}Num\ {\isadigit{0}}x{\isadigit{8}}{\isadigit{1}}{\isadigit{0}}{\isadigit{0}}{\isacharparenright}{\isacharparenright}\ {\isacharparenleft}Num\ {\isadigit{1}}{\isadigit{6}}{\isacharparenright}\ {\isacharparenleft}FieldFirst\ Type{\isacharunderscore}Length{\isacharunderscore}TPID{\isacharparenright}{\isacharcomma}\isanewline
\ \ Edge\ Type{\isacharunderscore}Length{\isacharunderscore}TPID\ TPID\ {\isacharparenleft}Eq\ {\isacharparenleft}FieldValue\ Type{\isacharunderscore}Length{\isacharunderscore}TPID{\isacharparenright}\ {\isacharparenleft}Num\ {\isadigit{0}}x{\isadigit{8}}{\isadigit{1}}{\isadigit{0}}{\isadigit{0}}{\isacharparenright}{\isacharparenright}\ {\isacharparenleft}Num\ {\isadigit{1}}{\isadigit{6}}{\isacharparenright}\ {\isacharparenleft}FieldFirst\ Type{\isacharunderscore}Length{\isacharunderscore}TPID{\isacharparenright}{\isacharcomma}\isanewline
\ \ \ldots\isanewline
{\isacharbrackright}{\isachardoublequoteclose}%
}